\documentclass[sn-mathphys-num]{sn-jnl}

\usepackage{graphicx}%
\usepackage{multirow}%
\usepackage{amsmath,amssymb,amsfonts}%
\usepackage{amsthm}%
\usepackage{mathrsfs}%
\usepackage[title]{appendix}%
\usepackage{xcolor}%
\usepackage{textcomp}%
\usepackage{manyfoot}%
\usepackage{booktabs}%
\usepackage{algorithm}%
\usepackage{algorithmicx}%
\usepackage{algpseudocode}%
\usepackage{listings}%

\usepackage{pifont} %
\usepackage{subfigure} %
\usepackage{lineno}  %
\usepackage{CJK} 

\theoremstyle{thmstyleone}%
\theoremstyle{thmstyletwo}%

\theoremstyle{thmstylethree}%

\begin{document}

\title{Detector description conversion and visualization in Unity for high energy physics experiments}


 \author[1]{\fnm{Tian-Zi} \sur{Song}}
 \author[1]{\fnm{Kai-Xuan} \sur{Huang}}
 \author[1]{\fnm{Yu-Jie} \sur{Zeng}}
 \author[1]{\fnm{Ming-Hua} \sur{Liao}}
 \author[2]{\fnm{Xue-Sen} \sur{Wang}}
 \author*[2]{\fnm{Yu-Mei} \sur{Zhang}}
 \email{zhangym26@mail.sysu.edu.cn}
 \author[1]{\fnm{Zheng-Yun} \sur{You}}

 \affil[1]{\orgdiv{School of Physics}, \orgname{Sun Yat-sen University}, \orgaddress{\city{Guangzhou}, \postcode{510275}, \country{China}}}

 \affil[2]{\orgdiv{Sino-French Institute of Nuclear Engineering and Technology}, \orgname{Sun Yat-sen University}, \orgaddress{\city{Zhuhai}, \postcode{519082}, \country{China}}}





\abstract{
While visualization plays a crucial role in high-energy physics (HEP) experiments, the existing detector description formats including Geant4, ROOT, GDML, and DD4hep face compatibility limitations with modern visualization platforms. 
This paper presents a universal interface that automatically converts these four kinds of detector descriptions into FBX, an industry standard 3D model format which can be seamlessly integrated into advanced visualization platforms like Unity. 
This method bridges the gap between HEP instrumental display frameworks and industrial-grade visualization ecosystems, enabling HEP experiments to harness rapid technological advancements. 
Furthermore, it lays the groundwork for the future development of additional HEP visualization applications, such as event display, virtual reality, and augmented reality.
}



\maketitle

\section{Introduction}

Visualization is indispensable at every stage in the life cycle of any high-energy physics~(HEP) experiment. 
From detector design, construction, assembly, and commissioning, to detector simulation, event reconstruction, data quality monitoring, event display, and physics analysis, visualization plays a crucial role in every part~\cite{BESIII_analysis}. 
Detector visualization, in particular, serves as the foundation for all other forms of visualization functions and holds an especially important position. 
In ``A Roadmap for HEP Software and Computing R\&D for the 2020s"~\cite{Roadmap}, it is highlighted that one of the main challenges in the field of HEP experiment visualization today is the sustainability of many experiment-specific visualization tools. 
In a rapidly evolving technological landscape, adopting industry-standard tools and common formats for visualization in HEP experiments offers a more cost-effective and accessible approach. 

In recent years, various detector geometry formats have been employed in the offline software of HEP experiments, such as Geant4~\cite{Geant4}, ROOT~\cite{ROOT}, Geometry Description Markup Language~(GDML)~\cite{GDML}, Detector Description Toolkit for High Energy Physics~(DD4hep)~\cite{dd4hep}, and GeoModel~\cite{GeoModel}. These geometry toolkits for detector descriptions can be used in Geant4. 

Filmbox~(FBX)~\cite{FBX},
a versatile file format created by Autodesk~\cite{autodesk}, supports 3D models and animations. 
Its comprehensive features and extensive compatibility make it a popular choice across diverse industries such as game development, film production, virtual reality~(VR)~\cite{VR} and augmented reality~(AR)~\cite{AR}. 
As an industry-standard format widely adopted across various sectors, FBX benefits from continuous support and development by Autodesk and the broad user community. 
This ensures its compatibility with emerging tools and technologies, making it a reliable choice for long-term, large-scale projects. 

Unity~\cite{unityweb, Unity} is a widely-used cross-platform game engine renowned for its powerful real-time 3D rendering and interactive capabilities. 
It serves as a powerful platform for 3D model editing, enabling the import and processing of FBX files, while also providing opportunities for the development of additional functionalities based on the framework. 
Leveraging this foundation, detector visualization outputs can be produced in various formats, such as software applications or HTML, enhancing the accessibility and dissemination of the visualization results. 

However, the four aforementioned detector description formats, Geant4, ROOT, GDML, and DD4hep, although commonly used in the HEP community, cannot be 
directly imported into Unity for visualization. 
Therefore, developing a unified interface that automatically converts these four detector formats into the FBX format will greatly benefit the HEP software community, enabling seamless integration, visualization and manipulation of detectors within Unity.

In this study, we propose a comprehensive method to automatically convert detector descriptions from all four formats into FBX format. 
For HEP detector descriptions already constructed in offline software, such as those from JUNO~\cite{JUNO}, EicC~\cite{EicC}, BESIII~\cite{BESIII}, and CEPC~\cite{CEPC} experiments, direct conversions to FBX format have been realized, demonstrating its feasibility by applications in different HEP experiments. 
Our method facilitates the seamless integration of detector descriptions from various HEP software with industrial modeling software, thereby establishing a robust foundation for advanced visualization efforts. 

The remainder of this paper is structured as follows. 
In Section~\ref{sec:detector_description}, we introduce the four commonly used detector descriptions in HEP offline software, FBX 3D modeling, and Unity. 
In Section~\ref{sec:methodologies}, the method of automatic conversion from the HEP detector description to FBX format is presented.
Section~\ref{sec:application} introduces the applications of this method in four HEP experiments and visualization of the detectors in Unity. 
Section~\ref{sec:advantages} discusses the advantages of the method and its further applications in HEP experiments. 
Finally, Section~\ref{sec:summary} provides the summary. 

\section{Detector description}
\label{sec:detector_description}

\subsection{Detector description in HEP software}

In the modern HEP software community, detector descriptions are commonly constructed using the following four software: Geant4, ROOT, GDML, and DD4hep. 

Geant4 is a widely-used software toolkit for simulating particle interactions with matter~\cite{Geant4}. 
It plays a critical role in the global HEP community, being essential for many experiments, including the ATLAS~\cite{atlas} and CMS~\cite{CMS} detectors at the Large Hadron Collider~(LHC)~\cite{LHC} and the neutrino experiments such as JUNO~\cite{JUNO}. 
Geant4's ability to accurately model complex particle interactions and detector geometries makes it an indispensable tool for experimental design, detector performance optimization, and data analysis. 
While Geant4 can handle basic visualization for detector and physics events, it requires different drivers and renderers. 
It supports a range of visualization drivers, such as OpenGL~\cite{openglweb}, VRML~\cite{VRML, VRMLweb, VRMLgeo}, and HepRep~\cite{heprep}, which must be configured to visualize detector geometries and particle trajectories. 
As a result, although Geant4 provides basic visualization tools, setting them up can be somewhat challenging, especially for users unfamiliar with the specific drivers and configuration options. 

ROOT is an open-source data analysis framework, extensively used in HEP as well as other scientific disciplines~\cite{ROOT}. 
It provides a robust suite of tools for data processing, statistical analysis, plotting, visualization, and efficient management of large-scale datasets, and it especially excels at handling and analyzing extensive data volumes. 
Its geometry module facilitates the 3D visualization of detector geometries, enabling detailed exploration and debugging of complex detector structures. 
ROOT has been utilized for detector description in 
detector design for the Electron-ion Collider in China~(EicC)~\cite{EicC}. 
Despite its strengths in visualizing physics data, such as through histogram plotting, the geometric visualization features in ROOT are limited in rendering quality and interactivity. 

GDML is an XML-based markup language designed for describing geometric structures, particularly used to define complex detector geometries in HEP experiments. 
Its primary function is to facilitate the storage and exchange of geometric data between different simulation platforms, serving as a standardized interface for geometry descriptions. 
For instance, detector geometries constructed in Geant4 and ROOT can be exported in GDML format and subsequently imported into other applications. 
Both Geant4 and ROOT support the import and export of GDML files~\cite{RootGeantGDML}. 
Additionally, GDML can be integrated into the offline software of HEP experiments, such as those used by the JUNO~\cite{JUNO_GDMLSW, root_junotao} and BESIII experiments~\cite{BESIII_geo}, enabling seamless access to detector geometries.

DD4hep is a comprehensive detector description framework designed to support all stages of HEP experiments. 
It features a modular design, enabling support for multiple geometry description formats, and integrates seamlessly with simulation tools like Geant4 and ROOT~\cite{dd4hep}. 
In both the HEP Software Foundation~(HSF) community white paper~\cite{HSF} and the R\&D roadmap for HEP software and computing in the 2020s~\cite{Roadmap}, DD4hep is recommended as the preferred detector description tool for the next-generation HEP experiments. 
Indeed, DD4hep has already been adopted by several future experiments, including ILC~\cite{ILC}, CLIC~\cite{CLIC}, FCC~\cite{FCC}, CEPC~\cite{CEPC}, and STCF~\cite{STCF}.

\subsection{FBX format and Unity}
\begin{figure}
    \centering
    \subfigure[FBX]{\includegraphics[width=0.4\linewidth]{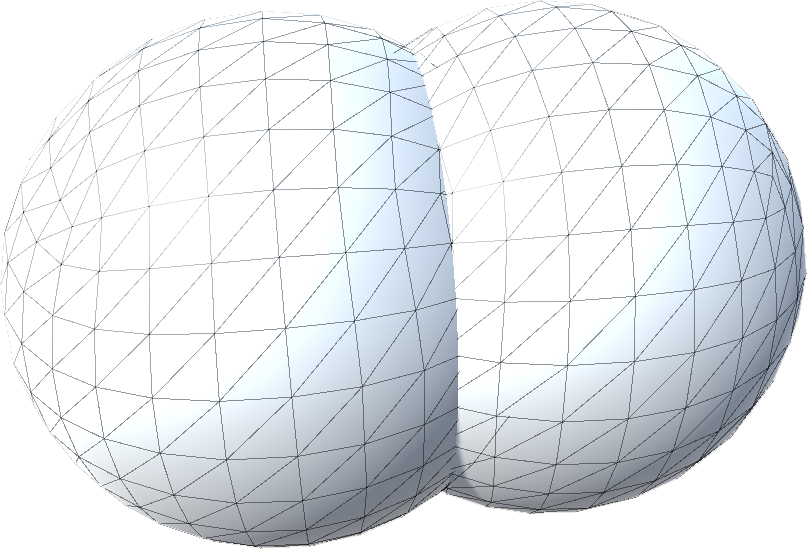}}
    \hspace{3pt}
    \subfigure[CSG]
    {\includegraphics[width=0.4\linewidth]{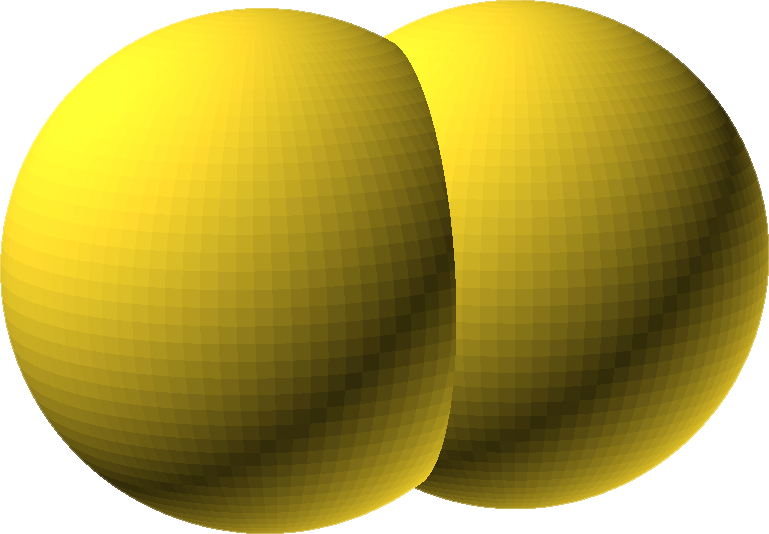}}
    \caption{Two spheres represented in FBX~(a) and CSG~(b) format. }
    \label{fig:fbx_mesh}
\end{figure}

FBX is a proprietary file format originally developed by Kaydara~\cite{FBX} and acquired by Autodesk in 2006~\cite{autodesk}. 
It utilizes 3D meshes to store and exchange 3D models, animations, and scenes, making it widely adopted in various industries. 
As opposed to formats like Constructed Solid Geometry~(CSG)~\cite{CSG} format, which is used to create solids in Geant4, the FBX format constructs geometric models through a mesh-based approach, relying on vertices, edges, and polygonal surfaces, which represents 3D models as solid entities. 
The use of mesh grids allows for intricate and detailed representations of complex geometries, which is particularly beneficial in applications requiring high fidelity in visualizations. 
Figure~\ref{fig:fbx_mesh} illustrates the surface structure of objects constructed using mesh representations in FBX and comparison with the CSG format.

A key advantage of FBX lies in its cross-platform compatibility, enabling seamless integration with major 3D software~\cite{FBX_all} such as 3D Studio Max~\cite{3dsmax}, Blender~\cite{blender}, Maya~\cite{maya}, Unity~\cite{fbx_unity}, and Unreal Engine~\cite{UnrealEngine}. 
This flexibility makes FBX a crucial exchange format, particularly in collaborative projects, cross-platform workflows, and scenarios where assets must be transferred across different software environments, offering a robust solution across various industries.

The structure of the FBX file format is optimized to efficiently handle large 3D models and complex scenes. 
The efficient representation of 3D models via meshes ensures high performance, which is crucial in game development, enhancing the efficiency of real-time rendering, especially in engines like Unity and Unreal, where FBX files are commonly used for scene importing and processing. 
Therefore, for the visualization of large and complex detectors in HEP experiments, FBX provides distinct advantages.

Unity is a powerful cross-platform game development engine developed by Unity Technologies~\cite{Unity}. 
While it is widely used in the gaming industry, it also has extensive applications in VR~\cite{VR}, AR~\cite{AR}, film production, architectural visualization, and automotive simulations. 
Some HEP projects have already successfully implemented visualizations using Unity, demonstrating its high feasibility for visualizing HEP experiment detectors, including ATLAS~\cite{ATLAS_AMELIA, ATLAS_AMLIA02}, BESIII~\cite{BESIII_unity}, BELLEII~\cite{Belle2_Unity} and JUNO~\cite{JUNO_unity, JUNO_liquid}. 
Overall, Unity offers several competitive advantages for HEP detector visualization, including: 

\begin{itemize}
    \item \emph{Cutting-edge 3D Display Capabilities.}
    Unity is a versatile platform for game development, ensuring long-term stability and continuous innovation. 
    Its visualization capabilities, real-time 3D rendering, interactive features, and ease of use consistently meet global industry standards.
    
    \item \emph{Cross-platform support.}
    Unity offers compatibility across multiple platforms, including MacOS, Linux, Windows, and web explorer. 
    This allows visualization results to be easily synchronized across different platforms, unlike the visualization tools within the HEP community, which are often limited to Linux environments. 
        
    \item \emph{Flexible Scripting.}
    Unity uses C\# for scripting, allowing developers to create complex interactions and customize engine functions to meet specific project requirements. 

    \item \emph{Long-term stability. }
    Unity is continuously supported by its developers with regular updates and new features, which ensures that projects built on Unity benefit from the latest advancements, offering long-term stability and making it a reliable choice for complex visualization tasks like HEP detectors with long life cycles and potential upgrades.
\end{itemize}


\section{Methodologies}
\label{sec:methodologies}
The HEP detectors usually consist of up to millions of detection units, forming highly complex structures that are difficult to rebuild manually in industrial 3D modeling software. 
Additionally, due to the need for efficient upgrades and future functionality development, HEP detector visualization requires a single source for detector description to maintain consistency between different applications. 
Without such consistency, the cost of maintaining visualization systems during detector upgrades would become prohibitive, compromising long-term usability. 
A unique source also ensures consistency in detector identification, facilitating the development of features like event displays, which rely on stable and consistent detector representations.

Fortunately, HEP detectors are typically constructed in simulation and offline software environments during their design phases, using software like GDML, ROOT, Geant4, or DD4hep. 
Therefore, we develop an automated interface to convert detector descriptions from any of these four popular formats in HEP software into the Unity-readable FBX format. 
By ensuring a single source of detector data, this approach enables the seamless transfer of different kinds of HEP detector descriptions into Unity for visualization while reducing long-term maintenance efforts and supporting further development of different functions.  

\subsection{Detector geometry conversion in HEP software}
\label{sec:hepgeo}
In HEP offline software, the requirement of geometry consistency across different applications has led to the development of several automated conversion interfaces~\cite{ROOT_Geant4, RootGeantGDML,dd4hep_simu}. 
These interfaces enable seamless format conversions between Geant4, ROOT, GDML, and DD4hep, ensuring consistency in detector descriptions across simulations~\cite{Lin:2022htc}, reconstructions~\cite{JUNO_reconstruct01, JUNO_reconstruct02, JUNO_reconstruc03}, event displays, and data analyses, which is crucial for HEP software development. 

For instance, in the BESIII experiment, the detector geometry is initially described using GDML. 
The GDML-Geant4 interface is employed for simulations, while the GDML-ROOT interface is used for reconstruction and event display~\cite{BESIII_geo, ROOT_Geant4, RootGeantGDML}. 
Additionally, a method for converting GDML detector descriptions into FBX format via a GDML-FreeCAD-Pixyz-FBX conversion chain has been proposed, allowing BESIII detector visualizations within Unity~\cite{BESIII_unity}. 
Furthermore, another approach has been proposed to directly convert DD4hep detector descriptions to FBX files, enabling visualization in FBX viewers or Unity~\cite{DD4hep_unity}. 
Recently, a method for importing detector geometry directly from GDML format into Unity~\cite{VR02} has been applied to the NICA Accelerator Consortium BM@N experiment at the Joint Institute for Nuclear Research.

These automated interfaces form a robust and reliable bridge between HEP detector descriptions and 3D modeling software. 
By ensuring a consistent and unique data source for detector geometry, they enable the seamless combination of advanced visualization capabilities. 
It not only minimizes manual labor for developers but also enhances the integration of sophisticated visualization tools, paving the way for future applications that demand higher precision and versatility in detector visualization. 

\subsection{Converting detector description to FBX format}
\begin{figure}
    \centering
    \includegraphics[width=0.7\linewidth]{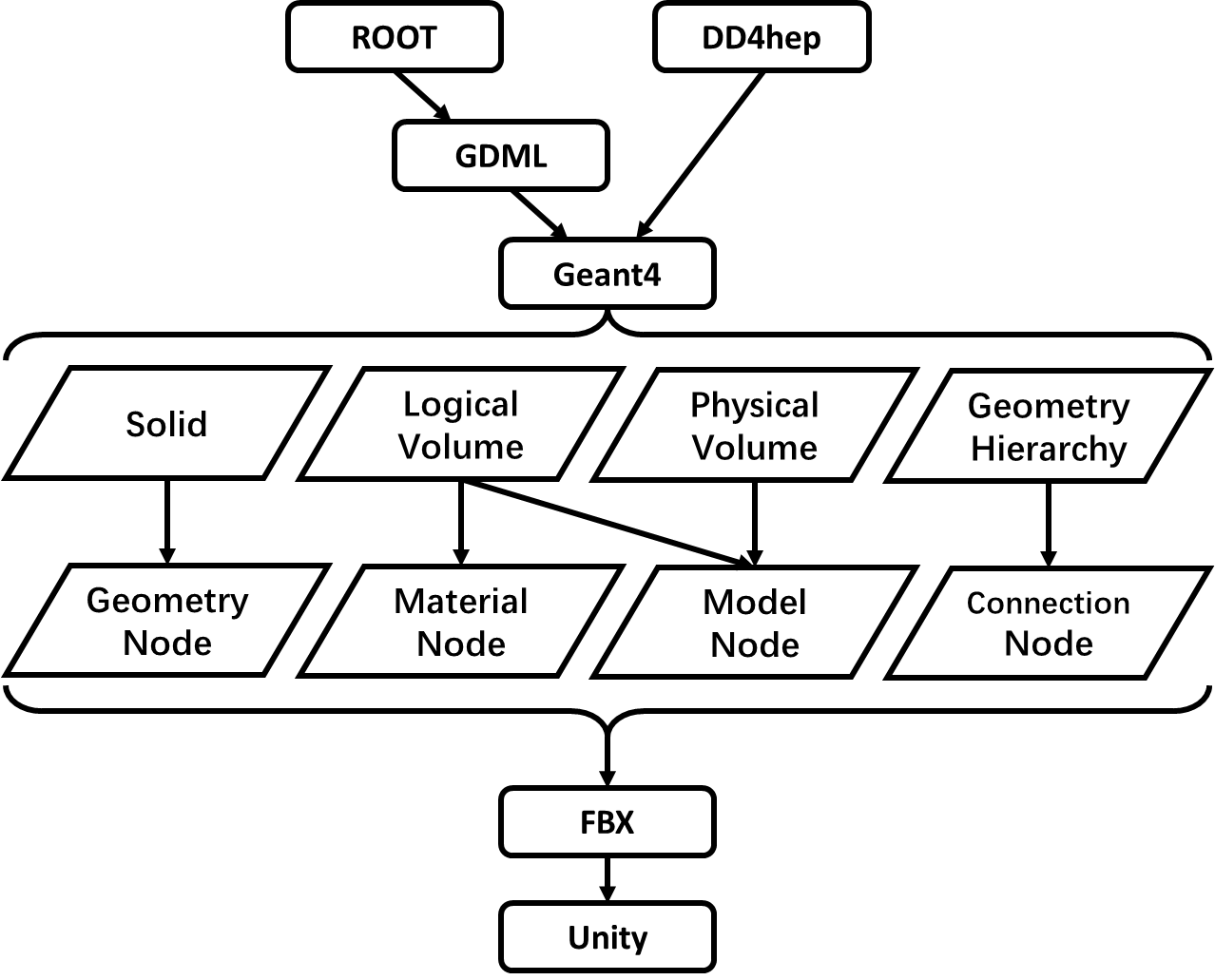}
    \caption{The data flow of detector description conversion from DD4hep, ROOT, GDML and Geant4 to FBX for 3D modeling in Unity.}
    \label{fig:liucheng}
\end{figure}

The data flow of detector description conversion from the formats in HEP offline software to the FBX format for 3D modeling in Unity is shown in Figure~\ref{fig:liucheng}. As mentioned in Section~\ref{sec:hepgeo}, there are already convenient interfaces for ROOT conversion to GDML and DD4hep conversion to Geant4. 
We can use these interfaces to construct detector in Geant4. 

In Geant4, the construction of 3D models is based on four key concepts: ``solid'', ``logical volume'', ``physical volume'', and the ``geometry hierarchy''. 
The solid element defines basic shapes and determines object dimensions, serving as the fundamental unit for model construction. 
The logical volume assigns material properties to the geometry. 
The physical volume establishes the spatial position, orientation, and management of multiple instances of the geometry. 
And the geometry hierarchy defines the structure of detector geometry, which organizes and constructs complex detector geometries in a tree-like structure. 
This system allows users to break down complex geometries into smaller, more tractable parts to manage. 
The whole process of building geometry in Geant4 fully reflects the application of the CSG system, which uses the basic geometric shapes (cube, sphere, cylinder, etc.) to construct complex geometric models through Boolean operations. 

In FBX, 3D models are represented using surface meshes and described with four main components: geometry node, material node, model node, and connection node, which correspond to geometry, material, model properties, and connections between them, respectively. 

\begin{figure}
    \centering
    \includegraphics[width=0.6\linewidth]{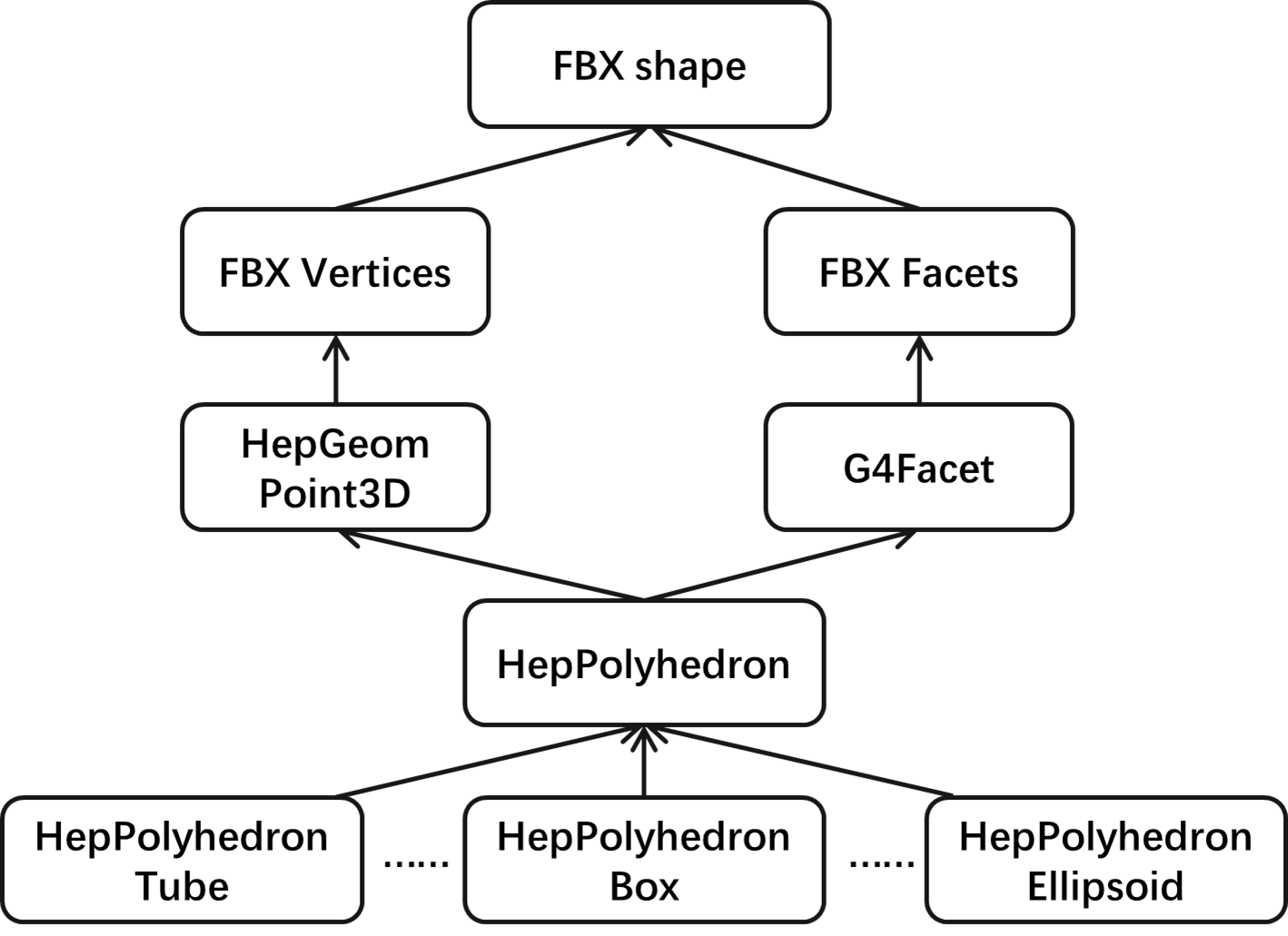}
    \caption{The conversion from the CSG shapes to Polyhedron in Geant4 and then to FBX shapes.}
    \label{fig:polygon}
\end{figure}

To preserve detector description integrity during the conversion from Geant4 to FBX, it is crucial to map the concepts in Geant4, including solid, logical Volume, physical volume, and geometry hierarchy, to their FBX counterparts, including geometry node, material node, model node, and connection node.
Due to the difference in solid definitions in the CSG and FBX format, a critical step is to automatically map the CSG solids into the solids defined with facets in FBX format. 
This mapping process is realized with the G4Polyhedron class in the Geant4 geometry package, which is detaily shown in Figure~\ref{fig:polygon}.  
G4Polyhedron is an intermediate class between Geant4 and visualization systems. 
It is intended to provide services for polygonization of the Geant4 shapes with triangulation and quadrangulation of complex polygons, and calculation of the normals for faces and vertices, which mirrors the solid definition in FBX format, ensuring geometric consistency for seamless visualization. 

An extension of geometry writer has already been developed in the HSF visualization framework~\cite{HSFWriter}, which has two packages to write the detector description from Geant4 into file for permanent persistency. 
One is the FBXWriter to export the Geant4 geometry into FBX file, and the other is the VRMLWriter to export the Geant4 geometry into VRML~\cite{VRML} file.
Then the converted FBX files can be imported into software such as Unity for 3D model display and rendering, while the VRML files can also be utilized for VR visualization. 
These two writers significantly enhance the visualization potential of HEP detectors and have been validated in BelleII~\cite{Belle2_Unity}.

\begin{figure}
    \centering
    \subfigure[One layer of MUC before fixing the bug.]{\includegraphics[width=0.4\linewidth]{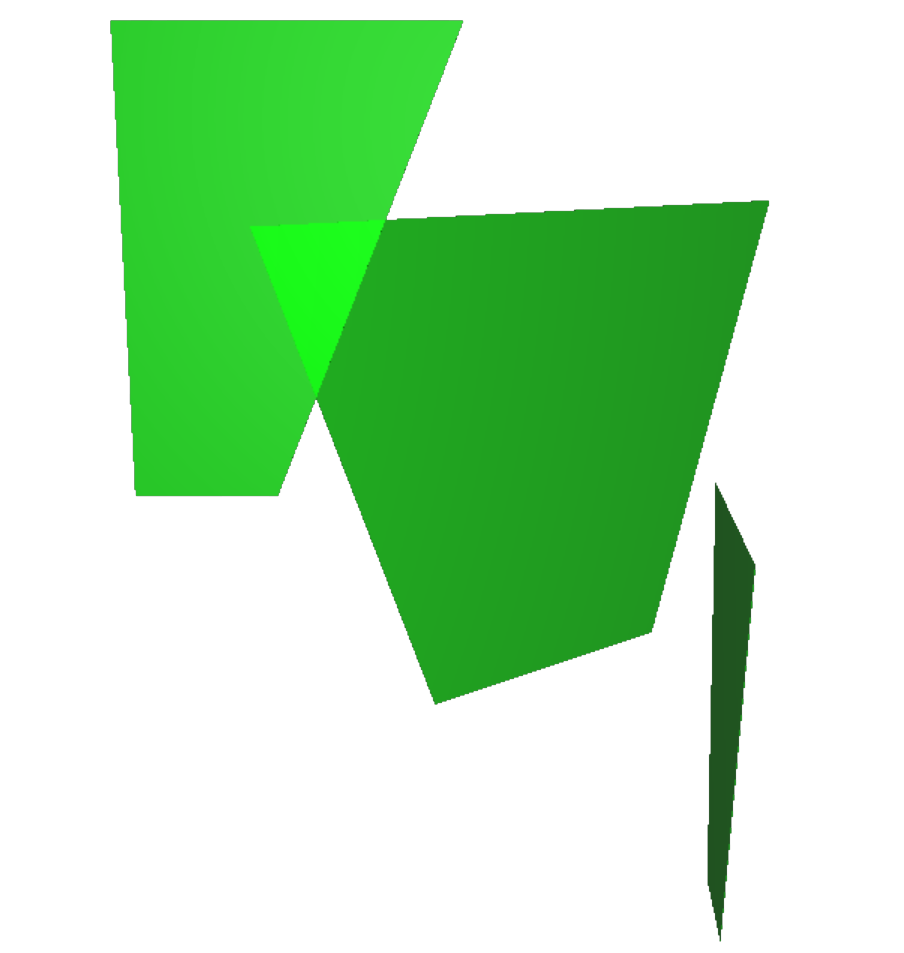}}
    \hspace{20pt}
    \subfigure[One layer of MUC after fixing the bug.]{\includegraphics[width=0.4\linewidth]{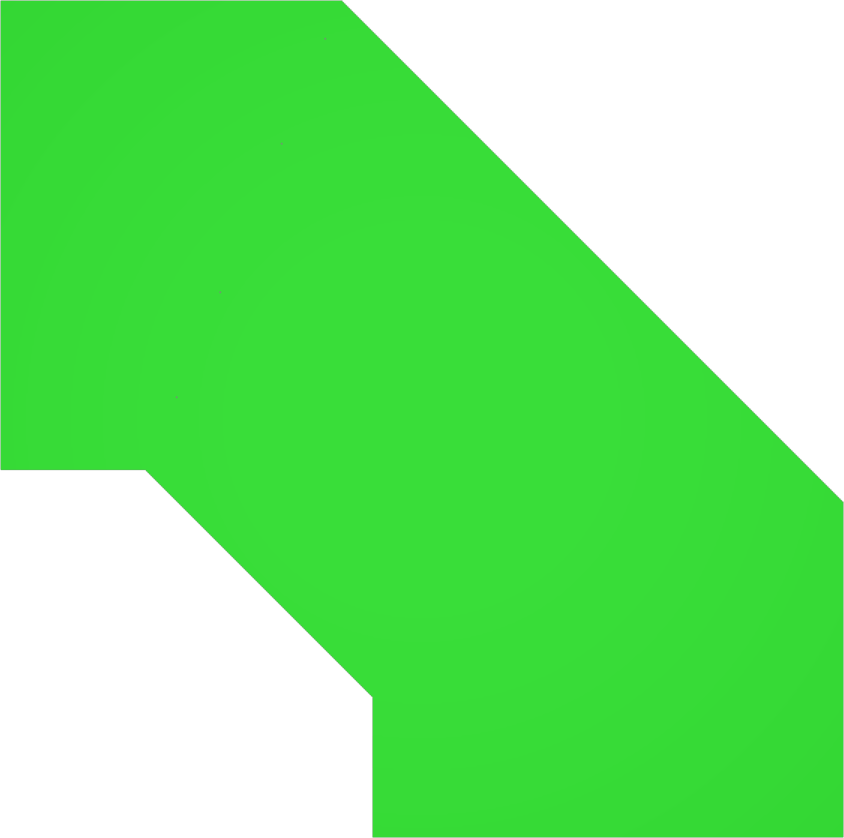}}
    \\
    \subfigure[The full MUC before fixing the bug.]{\includegraphics[width=0.41\linewidth]{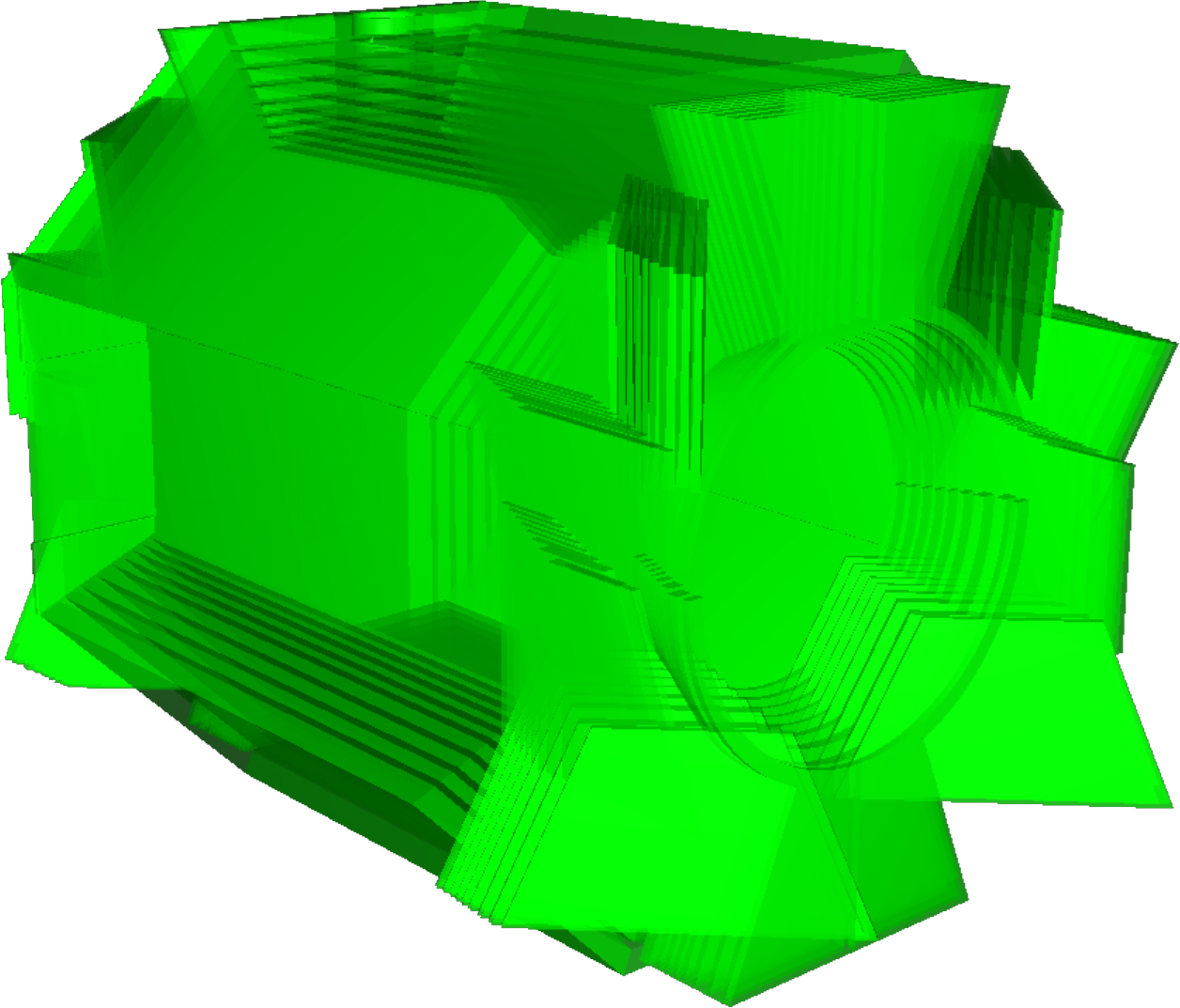}}
    \hspace{20pt}
    \subfigure[The full MUC after fixing the bug.]{\includegraphics[width=0.4\linewidth]{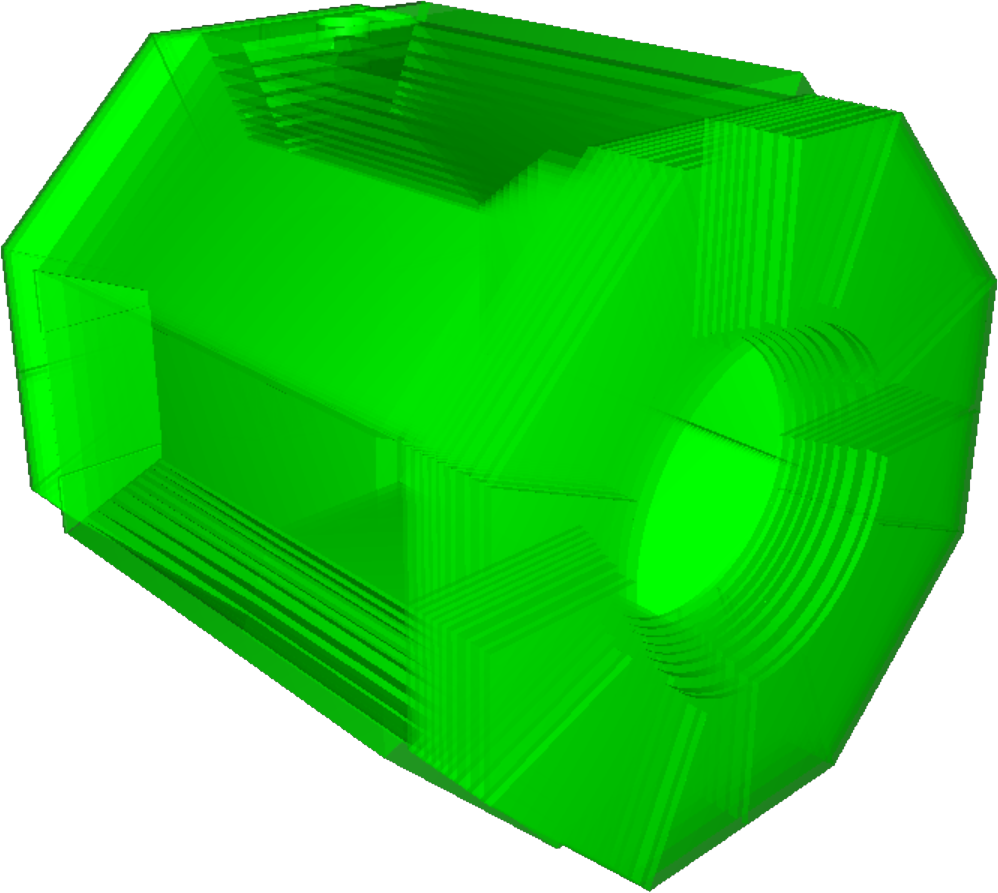}}
    \caption{Visualization of the BESIII Muon Counter end-cap before and after fixing a bug in GDML-Geant4 parser. The shapes are converted from GDML to FBX, and displayed in FBX Review.}
    \label{fig:Boolean}
\end{figure}

To ensure the accuracy of the geometric conversion process, it is imperative to validate the integrity of the geometry upon interfacing, both during input and output operations. 
Given that the interface operates within the Geant4 environment, certain adjustments to Geant4 Boolean operations are necessary to guarantee the correctness of the model. 
Complex geometric shapes are often misrepresented because Geant4 ignores the shape deviation caused by the sequence of rotation and position transformations in Boolean operations. 
Addressing these inaccuracies involves a straightforward correction of the rotational and translational order. 
As depicted in Figure~\ref{fig:Boolean}, before correcting the Boolean operations in Geant4, the converted FBX files exhibit geometric mismatch problems. 
In contrast, the images on the right exhibit the corrected transformation outcome after fixing the operations. 

The Muon Counter~(MUC) component of the BESIII detector achieves accurate visual representation solely after the above correction. 
Such meticulous adjustments ensure the fidelity and precision of the graphical display, both pre-conversion and post-conversion. 

The polygon edge count in circular approximations can be dynamically configured by adjusting the rotation step parameter during polyhedron generation. 
This parameter allows users to balance geometric fidelity against computational efficiency based on specific application requirements. 
By deviating from the default high-precision settings, the method effectively mitigates unnecessary computational overhead and storage expansion while maintaining adequate representation accuracy.

The source code of the FBX converter, originating from the HSF FBXWriter and updated with the features described above, is available in Github~\cite{mygithub} with an operational manual provided. 

\subsection{Visualization in Unity}
\label{subsec:UnityOperations}
After successfully converting the detector description to FBX format, it is essential to perform some basic settings in Unity to enhance its presentation quality. 
Currently, this paper identifies the following key settings:
\begin{itemize}
    \item \emph{Color of detector modules.}
    The detector loses its original color settings after conversion from GDML to FBX format. 
    Consequently, the models within the converted FBX files default to green, which can hinder observation and interaction, as shown in Figure~\ref{fig:Boolean}. 
    To enhance visualization, the colors of the detector modules can be redefined in Unity, as illustrated in Figure~\ref{fig:unitycode}. 
    By assigning distinct colors to differently named detector modules, we can effectively distinguish between various sub-detectors. This approach not only improves visual clarity but also provides a straightforward method for implementing the display of hits in event visualization.

\begin{figure}
    \centering
    \includegraphics[width=0.95\linewidth]{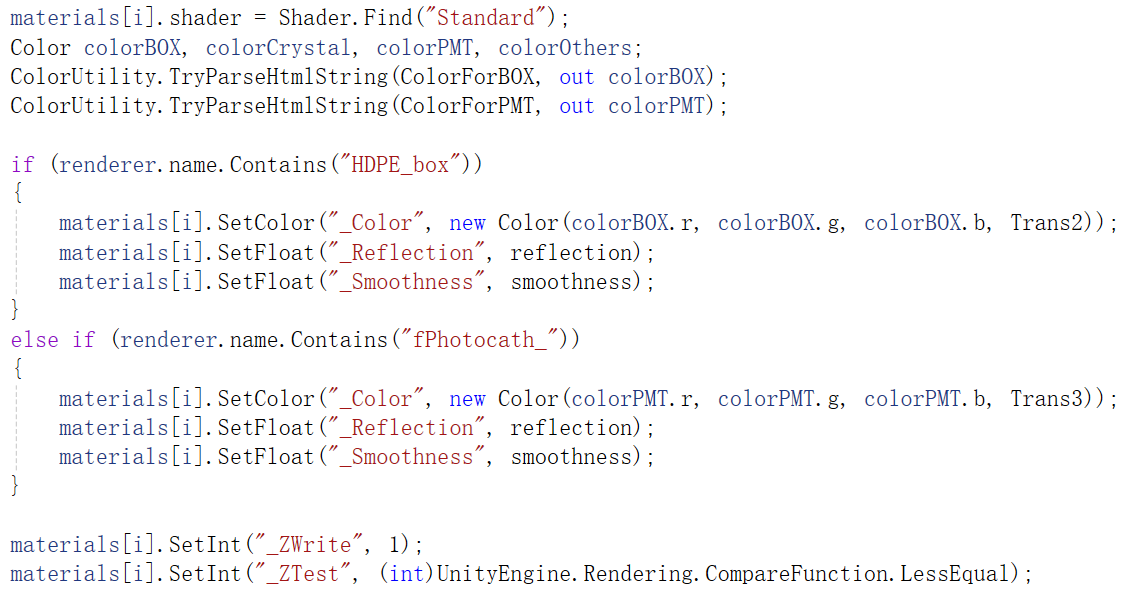}
    \caption{A piece of code in the Unity script to set the colors of the detector units. 
    }\label{fig:unitycode}
\end{figure}

    \item \emph{Transparency of detector units.}
    The transparency of certain detector units can hinder observation. 
    In Unity, the transparency can be adjusted through coding to enhance the visual display. 
    For instance, by increasing the transparency of the outer detector components, we can simultaneously observe both the inner and outer detectors. 
    This adjustment improves the overall clarity of the visualization, allowing for better analysis of interactions in the detectors.
    \item \emph{Other detector properties, light sources, and camera.}
    Parameters such as smoothness, material properties, texture, reflectivity, light source attributes, and camera functionalities can all be easily adjusted through script coding in Unity. 
    This flexibility allows for customized visual effects within the environment. 
    Particularly, modifications to camera angles and functionalities enable observations of the detector from various perspectives. 
    The subsequent sections of this paper will showcase the different display effects of various detectors, highlighting how perspective shifts can impact visualization outcomes.
\end{itemize}

\section{Applications}
\label{sec:application}
 
The comprehensive method of detector description conversion to FBX has been utilized in several HEP experiments. In this section, we will introduce its applications in four HEP detectors, each corresponding to one kind of detector description format, namely, Geant4, ROOT, GDML and DD4hep, and then visualize the detectors in Unity to demonstrate the feasibility of the method.  

\subsection{Geant4 to Unity with JUNO detector}

The Jiangmen Underground Neutrino Observatory~(JUNO) experiment, a 20 kilotons liquid scintillator detector in Guangdong, China, is primarily designed to determine the neutrino mass ordering~\cite{JUNO_old2}. 
The JUNO detector consists of three main components: the central detector~(CD), a surrounding water Cherenkov detector~(WP), and a top tracker~(TT)~\cite{JUNO}. 
The CD, housed in a stainless steel structure with photomultiplier tubes~(PMTs), is submerged in a water pool that functions as a cosmic muon veto. 
Above the water pool, the TT array accurately tracks muon trajectories, while a chimney connects the CD to an external calibration system, which is shielded from radioactivity. 

The JUNO simulation software~\cite{Lin:2022htc}, as well as its detector description, is based on the Geant4 toolkit. 
The Geant4 detector description can be seamlessly converted into FBX format using FBXWriter introduced in Section~\ref{sec:methodologies}, and the resulting FBX files can be imported into Unity for visualization. 

\begin{figure}
    \centering
    \subfigure[An overview of the JUNO detector. The topmost experimental hall is displayed in a semi-cylindrical shape, and beneath it are multiple layers of top trackers, all displayed in green. The outer grey cylinder below is the water pool, while the inner sphere is the central detector.]{\includegraphics[width=0.4\linewidth]{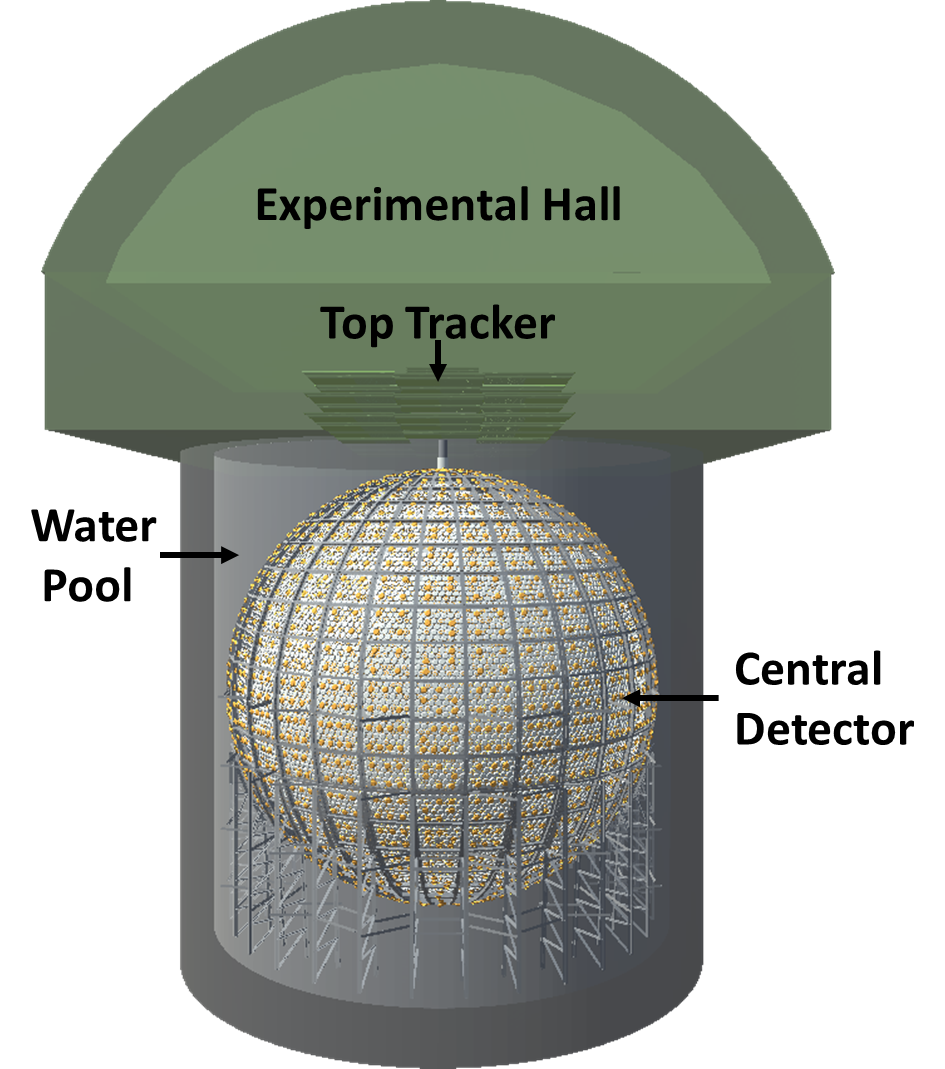}
    \label{fig:JUNOoverview}}
    \hspace{5pt}
    \subfigure[The JUNO Central Detector with PMTs supported by the stainless steel structure. A life-sized human model serves as a scale reference. The grey ribbons are the supporting structure, within which there are the semi-transparent acrylic ball and numerous PMTs displayed in yellow. ]{\includegraphics[width=0.55\linewidth]{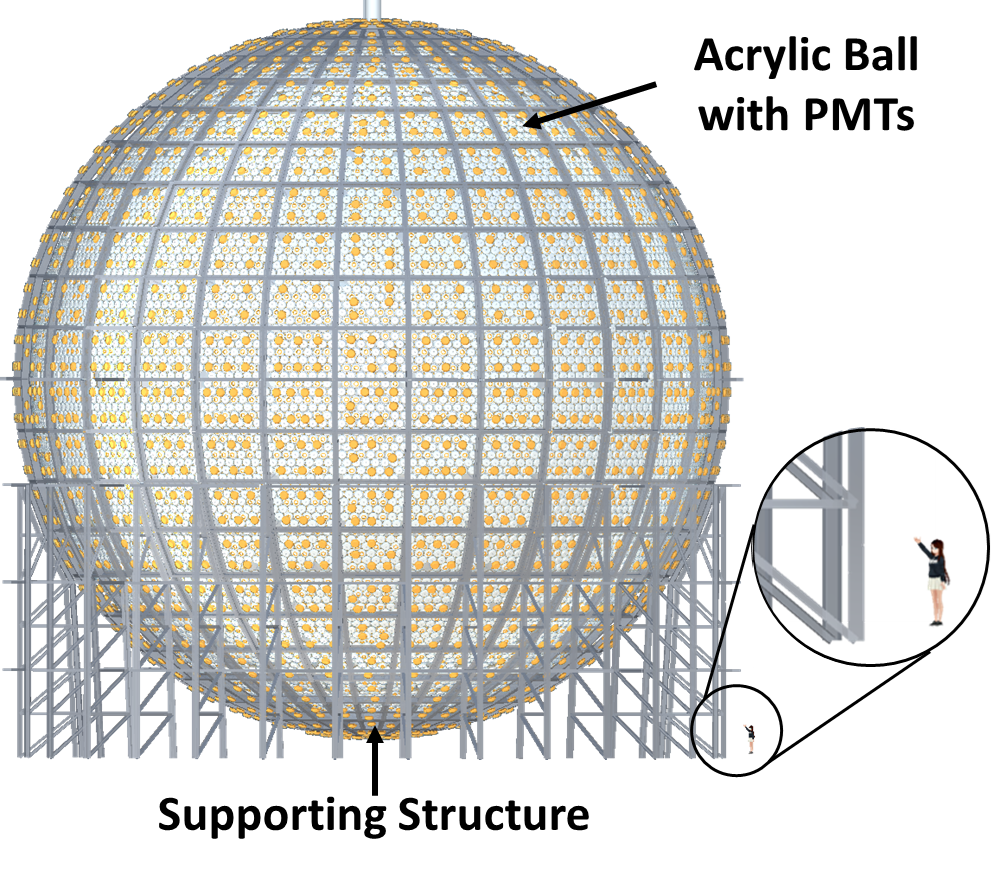}
    \label{fig:JUNOpeople}}  \\
    \subfigure[A close-up rendering showing part of the JUNO CD, including the stainless steel structure~(grey), the internal and external PMTs~(yellow), and the acrylic ball~(semi-transparent) containing 20,000 tons of liquid scintillator.]{\includegraphics[width=0.9\linewidth]{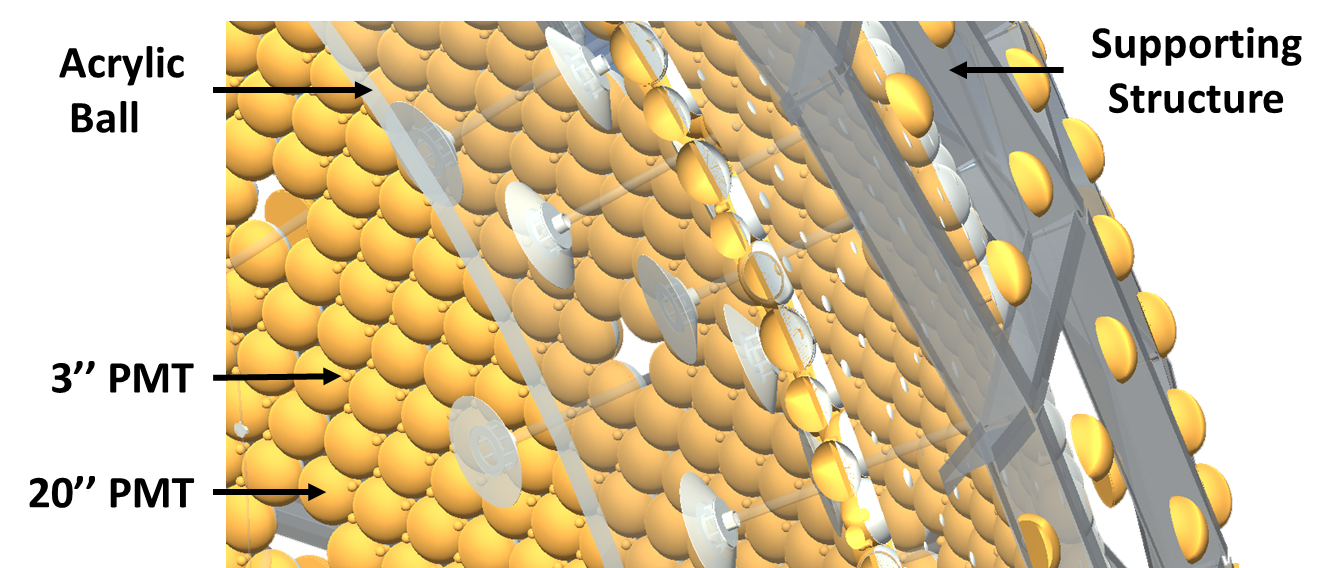}} 
    \caption{Visualization of the JUNO detector with detector description converted from Geant4 to FBX and displayed in Unity.}   
    \label{fig:JUNO}
\end{figure}


As illustrated in Figure~\ref{fig:JUNO}, Unity's flexibility allows for versatile viewing of the JUNO detector from various perspectives, including the overview of the JUNO 3D representation, the partial view highlighting the main acrylic ball with PMTs and its supporting structure, and even detailed close-ups. 
A life-sized human model is displayed on the right as a reference, which provides a clear understanding of the detector scale. 
This process is facilitated by Unity, allowing us to import relevant 3D characters directly from its extensive library at no cost, such as the girl model depicted in Figure~\ref{fig:JUNOpeople}. 

\subsection{ROOT to Unity with EicC detector}

The Electron-Ion Collider in China~(EicC) is a proposed facility aimed at studying the fundamental structure of matter by colliding electrons with protons and ions~\cite{EicC}. 
With energies of 3.5 GeV for electrons and 10-25
GeV for protons and Helium-3 nuclei, EicC focuses on exploring Quantum Chromodynamics~(QCD) phenomena, including the distribution of partons~(quarks and gluons) inside hadrons, the behavior of sea quarks, and the near-threshold production of heavy quarkonium. 

\begin{figure}
    \centering
    \includegraphics[width=0.8\linewidth]{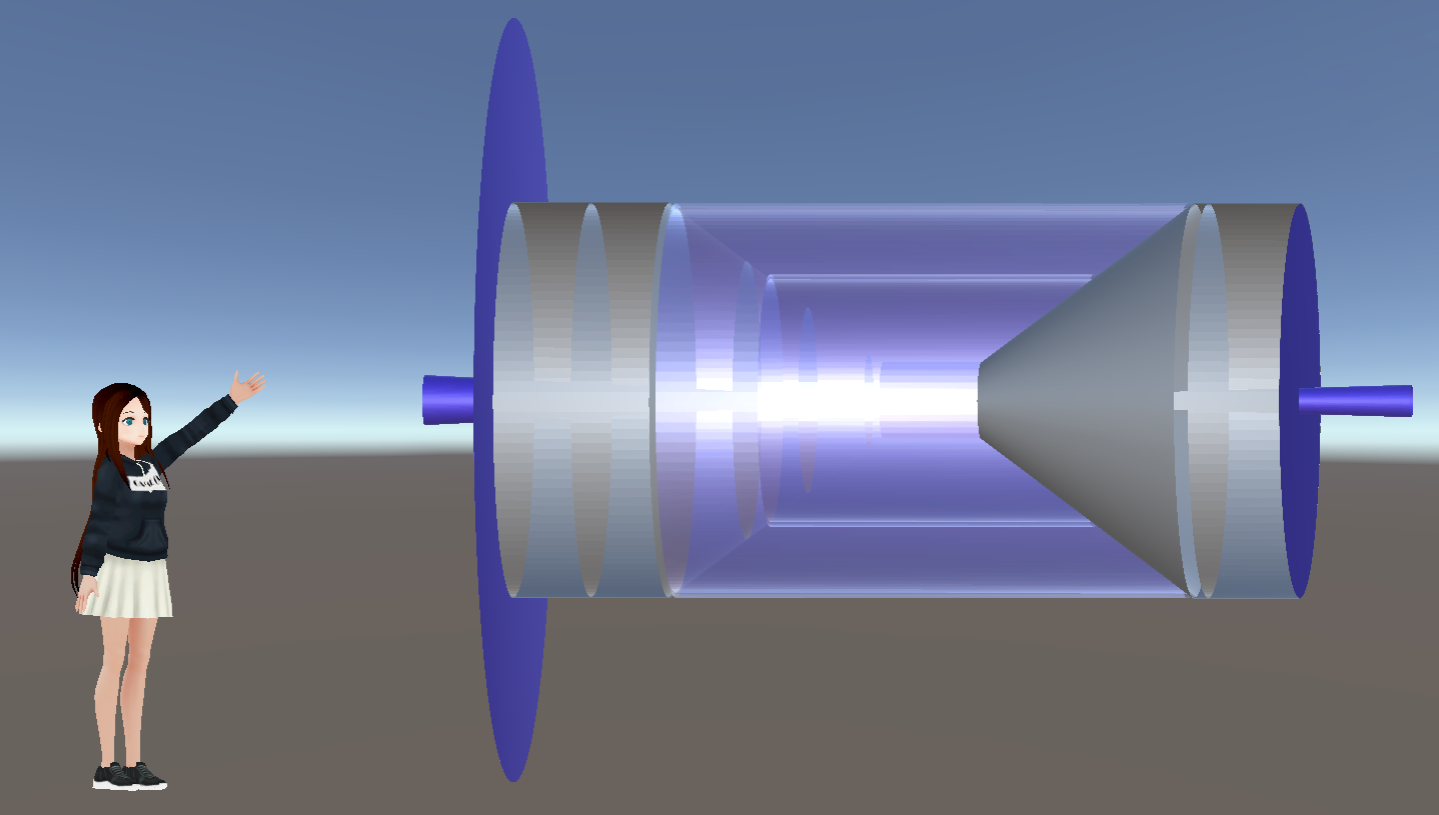}
    \caption{Display of the design of EicC detector converted from ROOT to Unity, featuring a life-sized human model on the left serves as a scale reference. The purple translucent barrel is the coil, and the darker cone inside is the Cherenkov detector.}
    \label{fig:EicC}
\end{figure}

In the current design of EicC, the detector description is constructed based on ROOT and can be exported as a GDML file using the ROOT geometry package. 
With our method, the GDML file can be converted into FBX files and imported into Unity for display. 
As shown in Figure~\ref{fig:EicC}, the purple translucent barrel is the coil, and the darker cone inside is the Cherenkov detector. 
Since the electron beam typically has much lower energy than the proton or ion beams, more particles are scattered in the forward direction, particularly on the side of the ion beam. 
As a result, a more dense and specialized arrangement of detectors is necessary in the forward region to capture these high-energy particles effectively. 


\subsection{GDML to Unity with BESIII detector}

The Beijing Spectrometer~(BESIII) detector~\cite{BESIII} is designed to record collisions from the Beijing Electron Positron Collider storage ring, operating within a center-of-mass energy range of 1.84 to 4.95 GeV.  
The detector geometry of BESIII, including the four sub-detectors, is described with GDML~\cite{ROOT_Geant4, BESIII_geo}. 
Its cylindrical core covers 93\% of the total solid angle and comprises a helium-based multilayer drift chamber~(MDC), a plastic scintillator time-of-flight system~(TOF), and a CsI(Tl) electromagnetic calorimeter~(EMC), all encapsulated within a superconducting magnet that provides a magnetic field of 1.0 Tesla. 
The outermost layer of the BESIII detector is comprised of the Multi-Layer Resistive Plate Chamber~(RPC) sub-detector, referred to as the MUC. 

\begin{figure}
    \centering
    \includegraphics[width=0.7\linewidth]{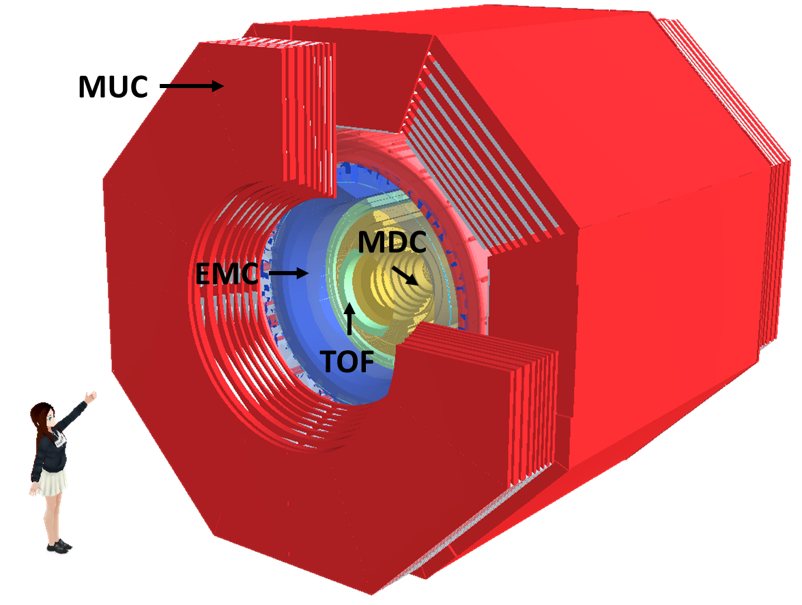}\\
    \caption{Display of the BESIII detector in Unity.  The yellow part is MDC, the green part is TOF, the blue part is EMC, and the outermost red part interleaved with white layers is MUC, where a quarter of the end cap is omitted for clarity. A life-sized human model on the left serves as a scale reference.}
    \label{fig:BESIII_whole}
\end{figure}

\begin{figure}
    \centering
    \subfigure[]{\includegraphics[width=0.42\linewidth]{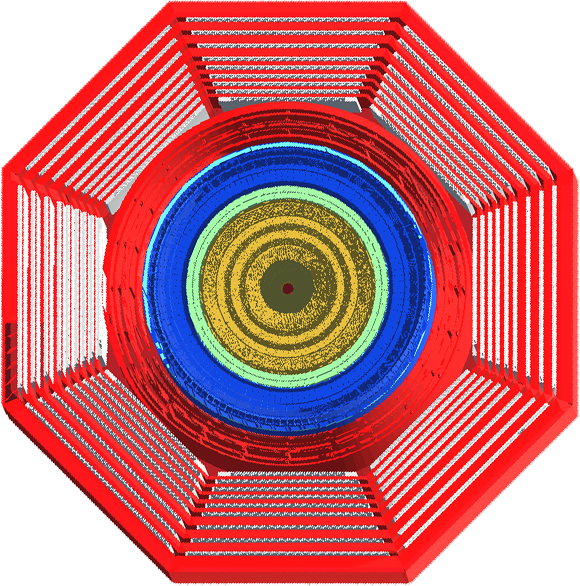}}
    \hspace{10pt}
    \subfigure[]{\includegraphics[width=0.47\linewidth]{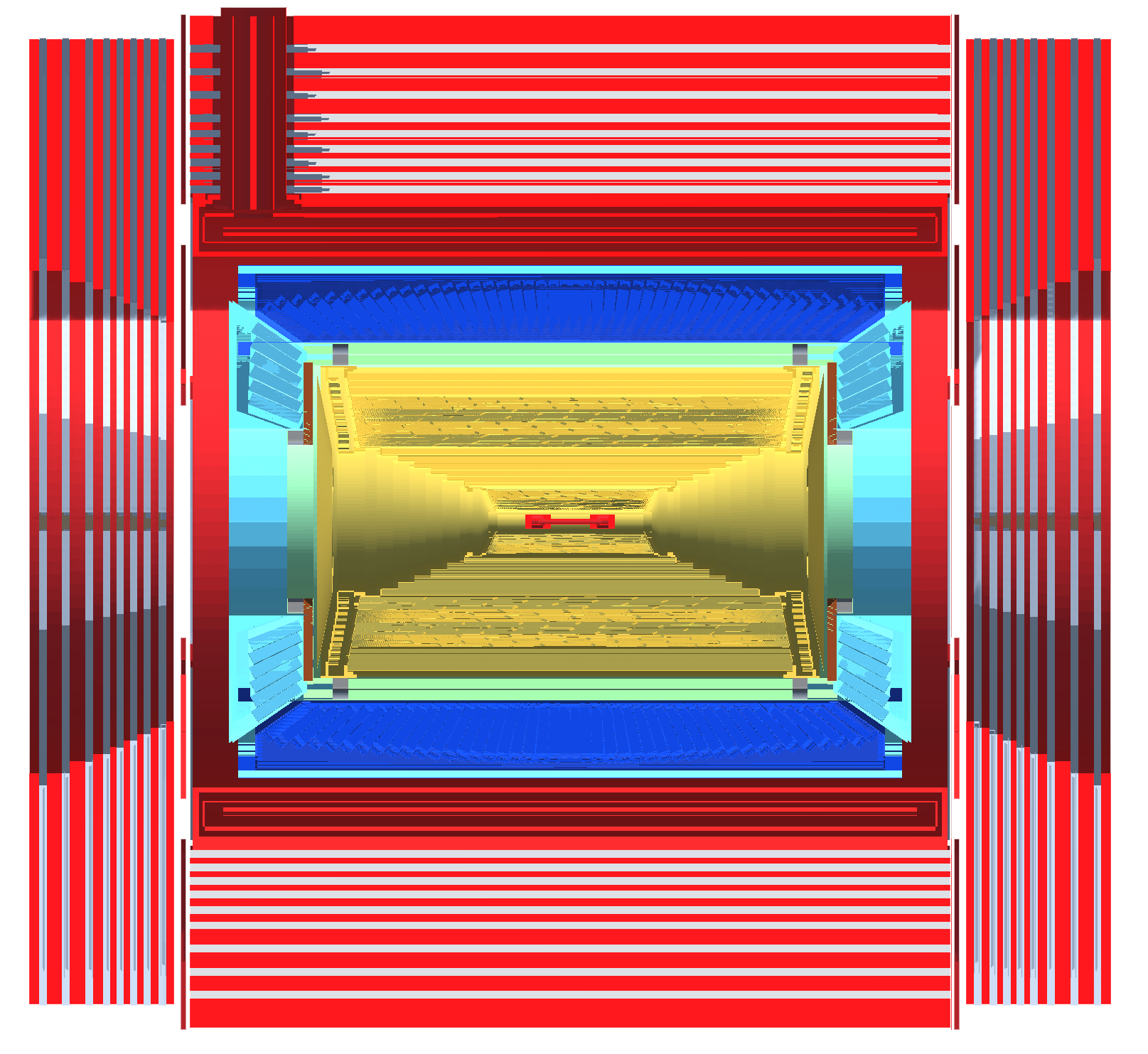}}  \\
    \caption{The X-Y cross-section view~(a) and Z-R view~(b) of the BESIII detector in Unity. }
    \label{fig:BESIII_cross}
\end{figure}

\begin{figure}
    \centering
    \subfigure[MUC]{\includegraphics[width=0.3\linewidth]{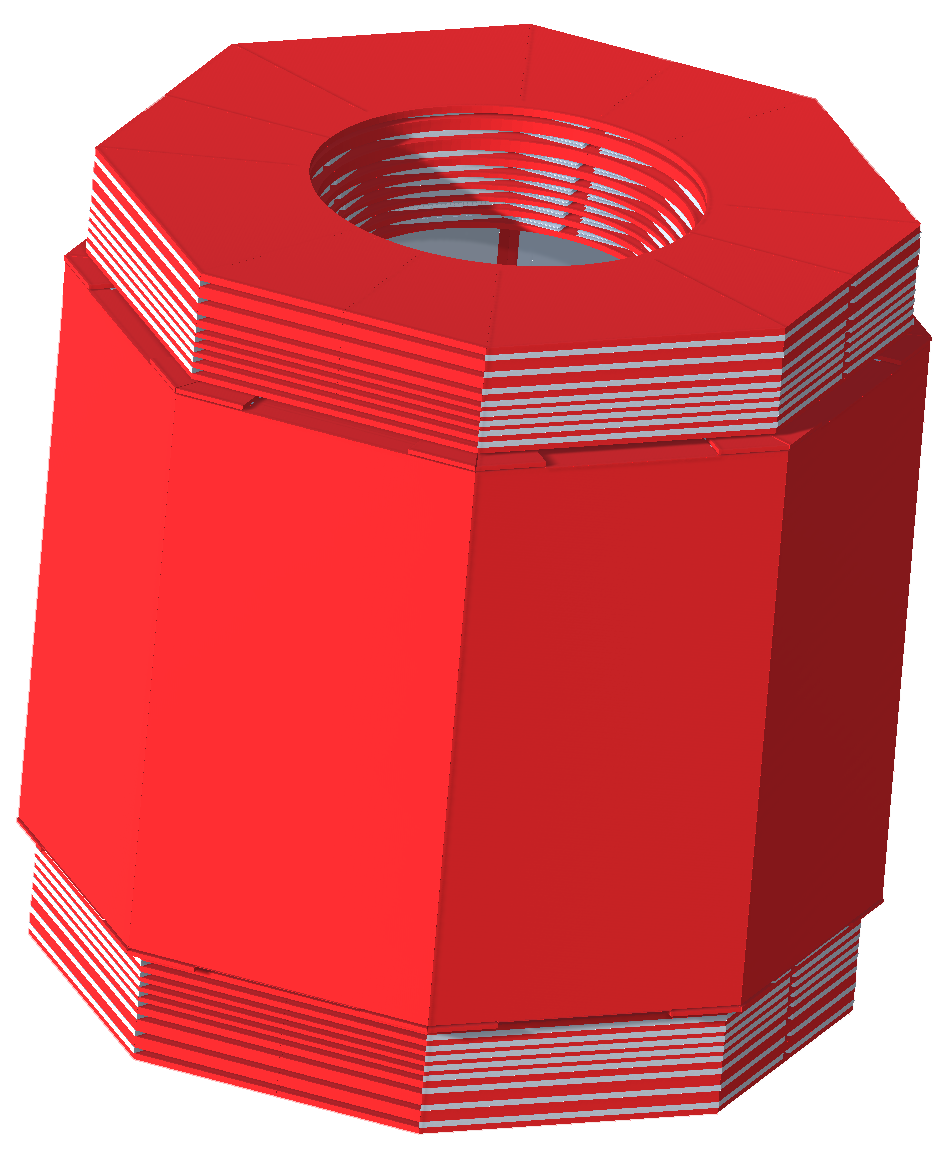}}
    \hspace{5pt}
    \subfigure[EMC]{\includegraphics[width=0.22\linewidth]{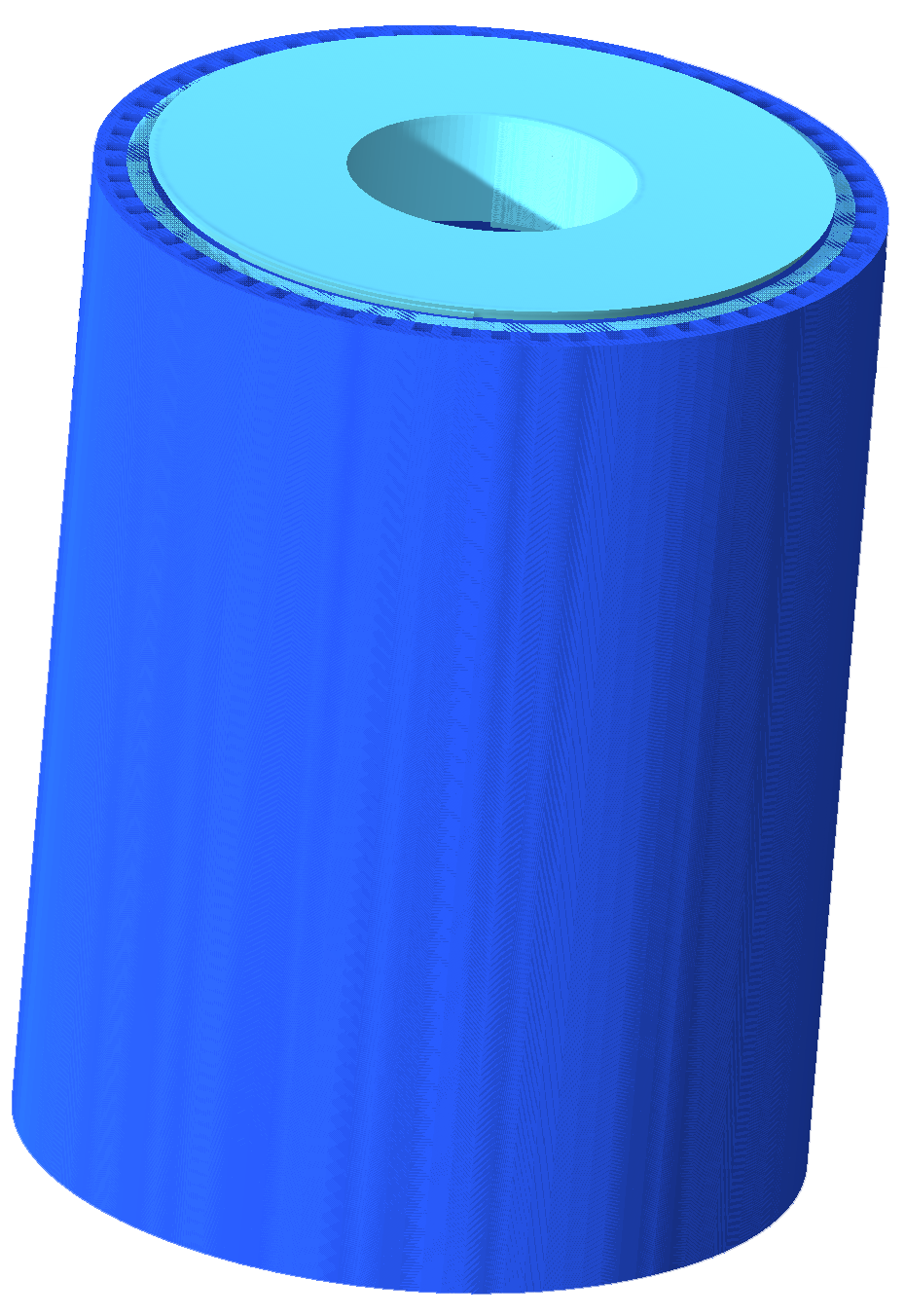}}
    \hspace{5pt}
    \subfigure[TOF]{\includegraphics[width=0.19\linewidth]{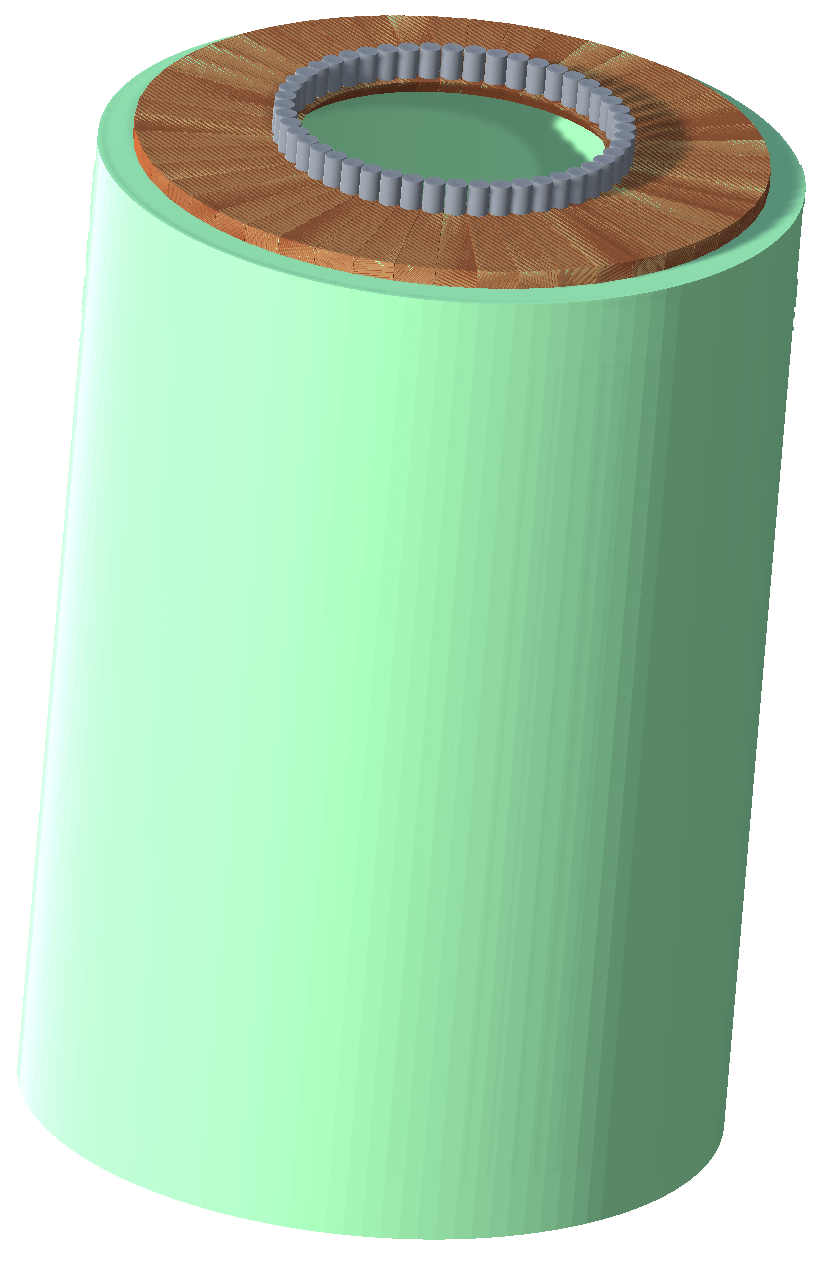}}
    \hspace{5pt}
    \subfigure[MDC]{\includegraphics[width=0.18\linewidth]{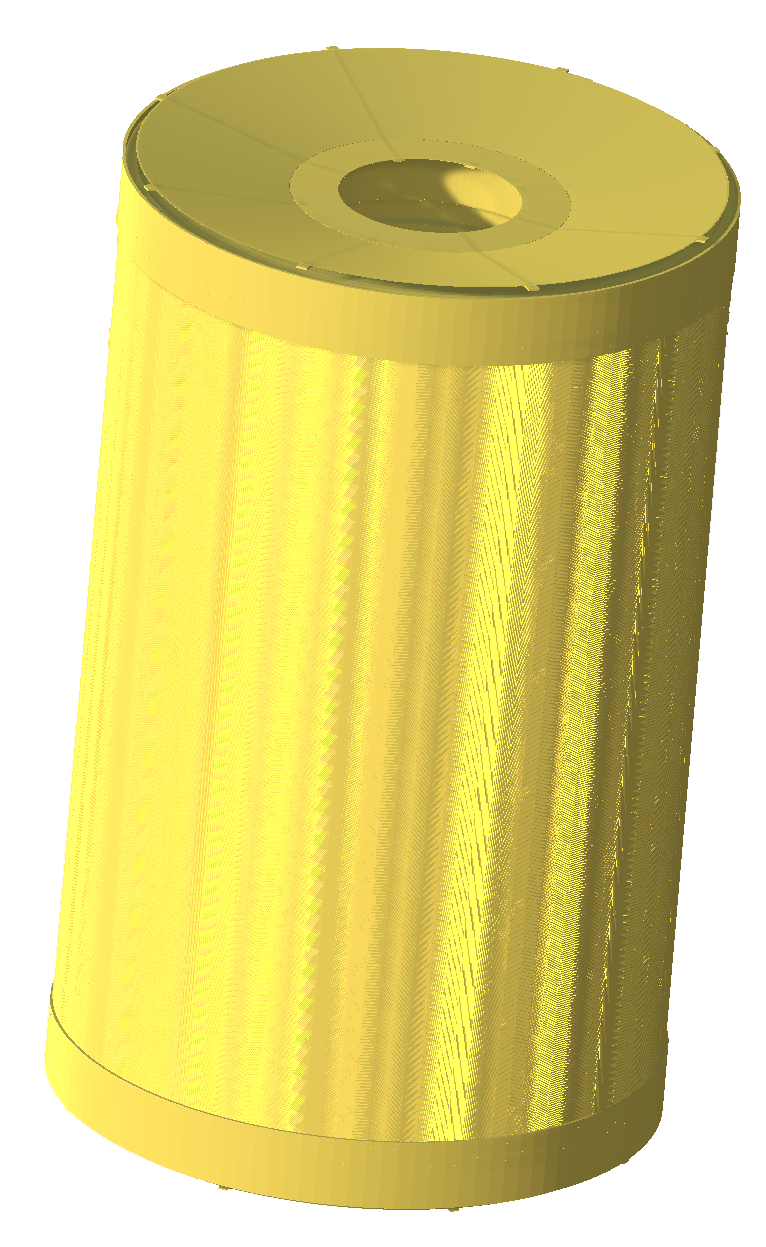}}  \\
    \caption{The sub-detectors of BESIII shown in Unity. Subfigures (a) to (d) follow the detector's geometrical layout from the outermost to innermost components. }
    \label{fig:BESIII_sub}
\end{figure}

The conversion from GDML to FBX format uses the GDML-Geant4 interface to construct detector in Geant4 as a bridge for FBX output described in Section~\ref{sec:hepgeo}. 
However, since an executable program combing the GDML-Geant4 interface and FBXWriter can be build, it allows easy conversion by simply running the program with the GDML file as input to export the FBX file.
In Unity, the detector can be viewed as a whole (with part of the MUC detector omitted for clarity), the X-Y cross-section view, the Z-R view, and easily display each sub-detector individually, as shown in Figure~\ref{fig:BESIII_whole}, Figure~\ref{fig:BESIII_cross} and Figure~\ref{fig:BESIII_sub}, respectively.

\subsection{DD4hep to Unity with CEPC detector}

The Circular Electron Positron Collider~(CEPC) is an international scientific facility proposed by the Chinese particle physics community and is currently in the R\&D phase~\cite{CEPC}. 
It is expected to produce nearly one trillion $Z$ bosons, 100 million $W$ bosons, and over one million Higgs bosons. 

As a next-generation high-energy collider experiment, the CEPC detector design is realized with the DD4hep software, with its detector description provided with DD4hep. 
After the detector construction in Geant4 with DD4hep as input, the detector description can also be exported to the FBX format. 

\begin{figure}
    \centering
    \subfigure[]{\includegraphics[width=0.45\linewidth]{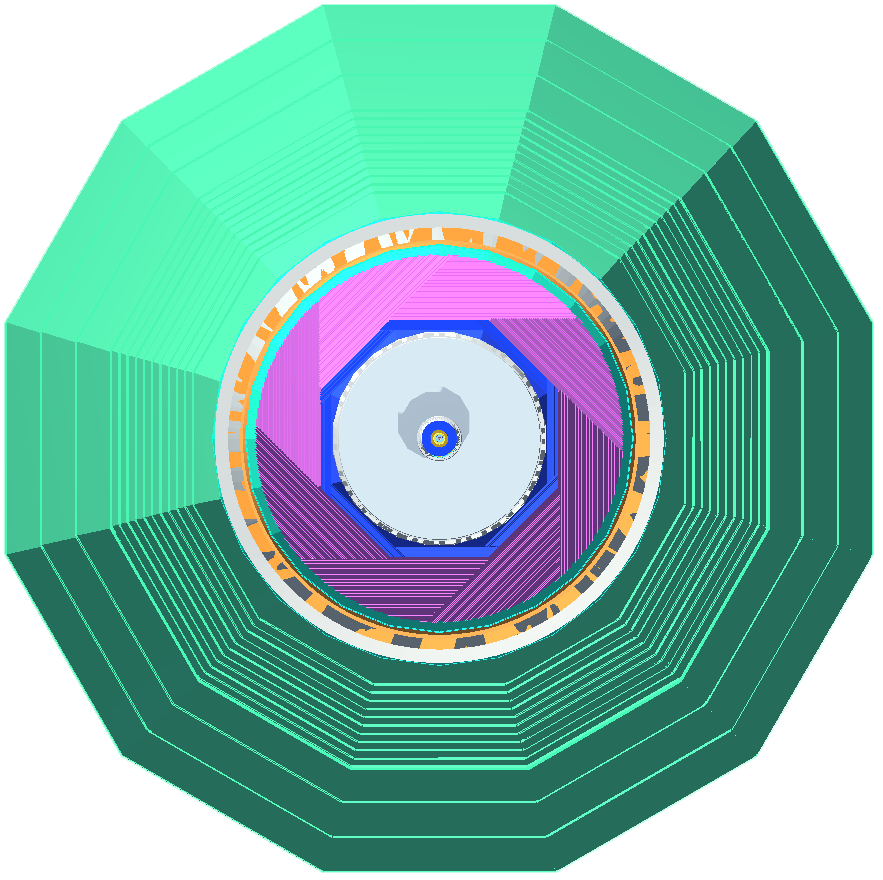}}
    \hspace{10pt}
    \subfigure[]{\includegraphics[width=0.40\linewidth]{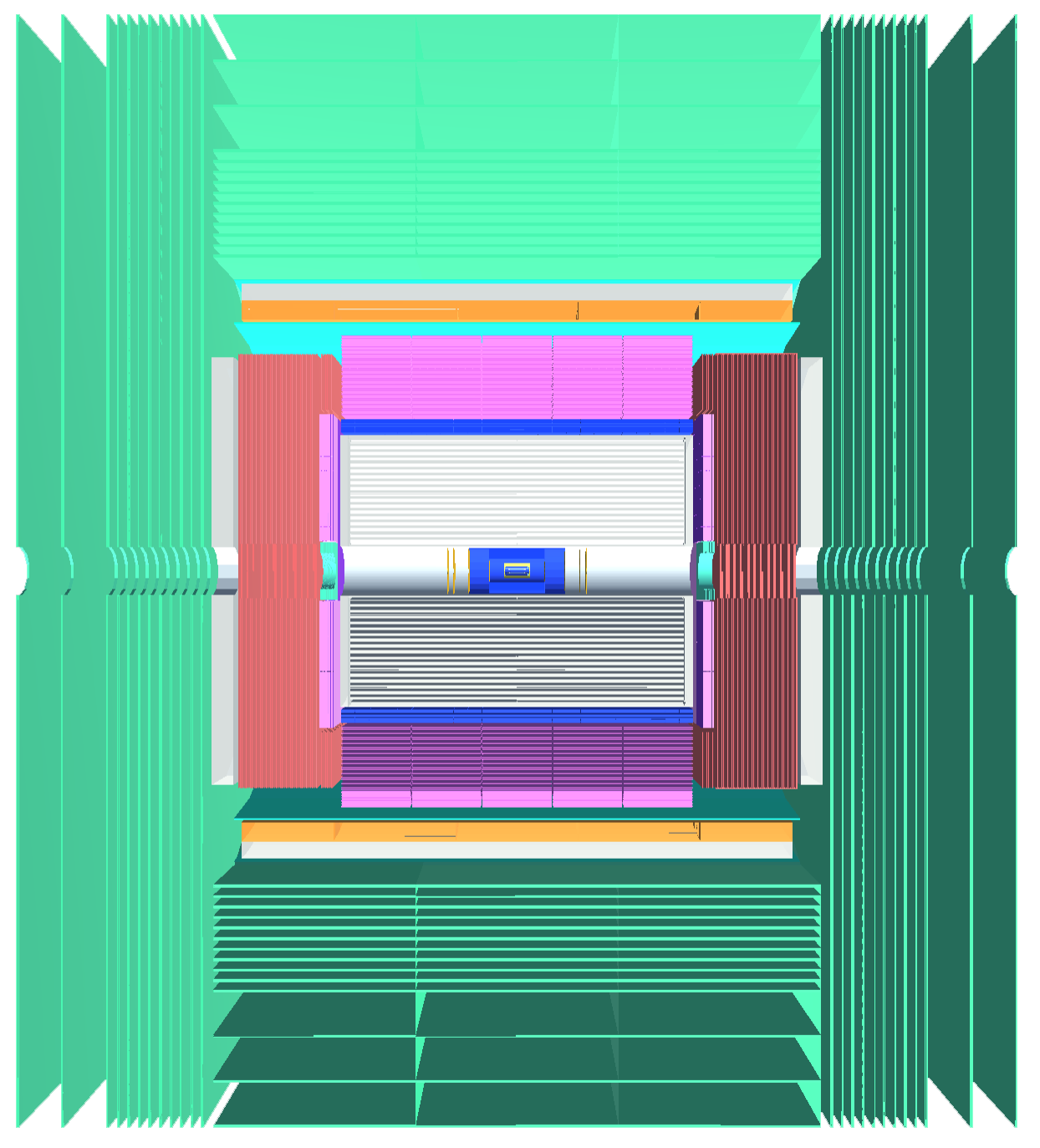}}
    \caption{The X-Y cross-section view~(a) and Z-R view~(b) of the CEPC detector in Unity. }
    \label{fig:CEPC_cross}
\end{figure}

As shown in Figure~\ref{fig:CEPC_cross}, from innermost to outermost, the CEPC detector comprises a silicon pixel vertex detector~(VTX), a Silicon Inner Tracker~(SIT), a Time Projection Chamber~(TPC) surrounded by a Silicon External Tracker~(SET), a silicon-tungsten sampling Electromagnetic Calorimeter~(ECAL), and a steel-Glass Resistive Plate Chambers~(GRPC) sampling Hadronic Calorimeter~(HCAL). 
Additionally, a 3 Tesla superconducting solenoid is integrated, along with a flux return yoke embedded with a Muon Detector~(MUD). 

\begin{figure}
    \centering
    \subfigure[MUD]{\includegraphics[width=0.3\linewidth]{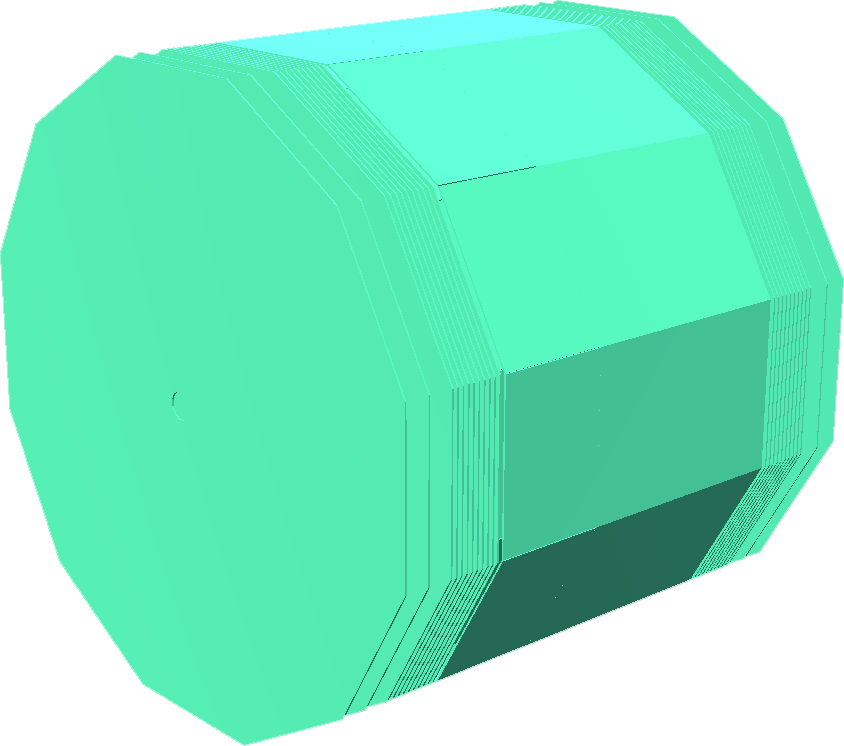}}
    \hspace{10pt}
    \subfigure[3 Tesla superconducting solenoid]{\includegraphics[width=0.29\linewidth]{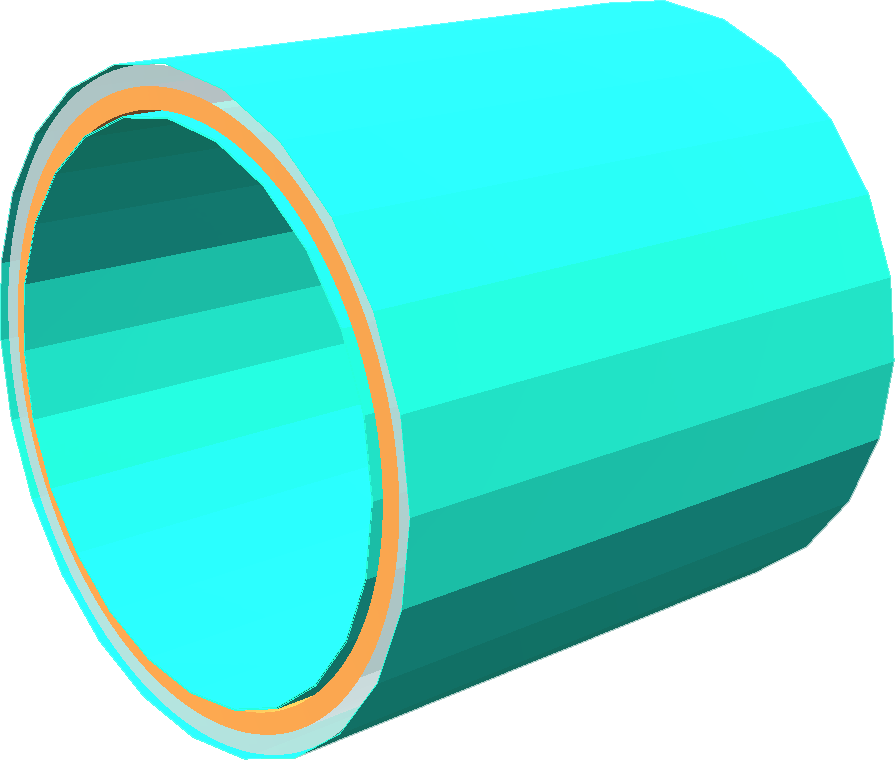}}
    \hspace{10pt}
    \subfigure[HCAL]{\includegraphics[width=0.3\linewidth]{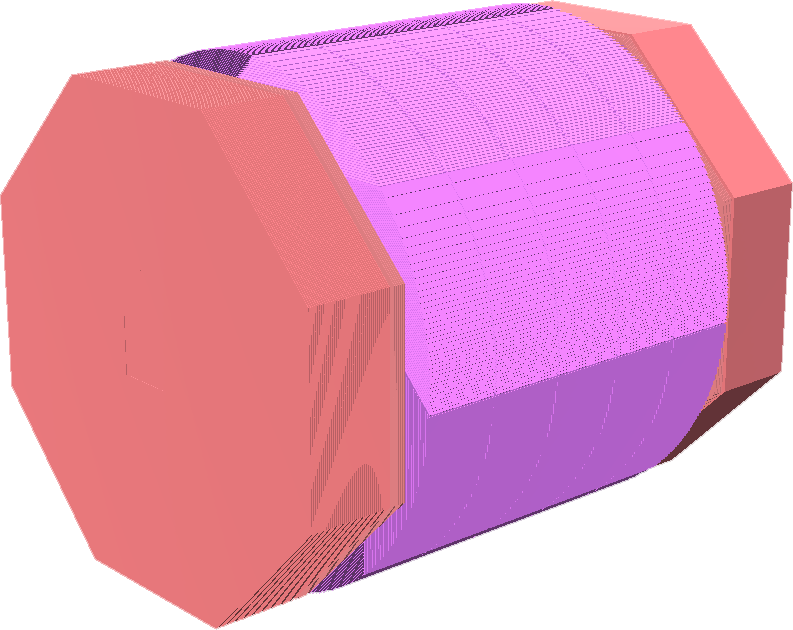}}\\
    \subfigure[ECAL]{\includegraphics[width=0.3\linewidth]{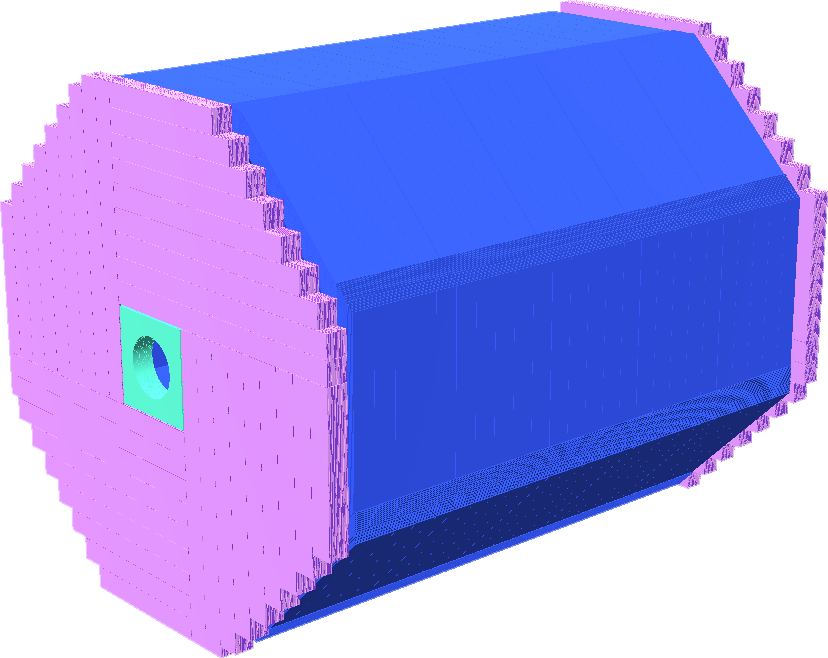}} 
    \hspace{10pt}
    \subfigure[SET, TPC \& SIT]{\includegraphics[width=0.3\linewidth]{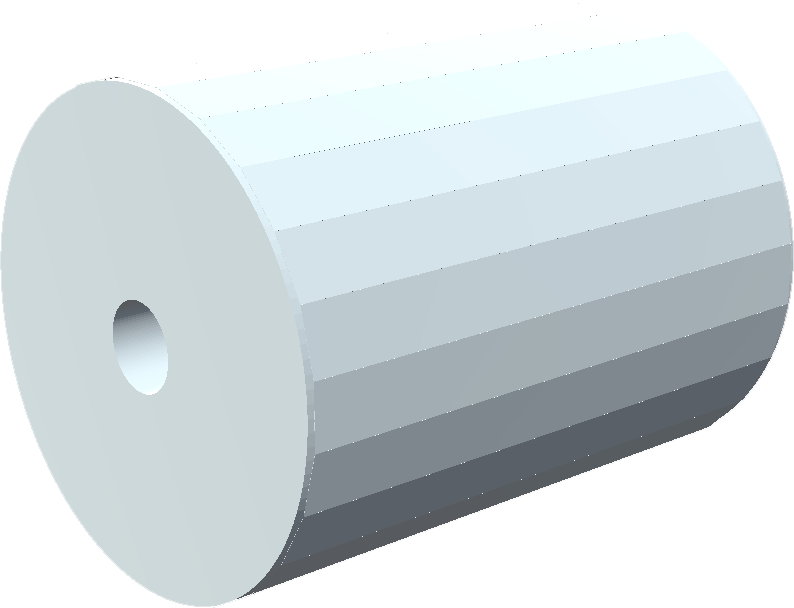}}
    \hspace{10pt}
    \subfigure[VTX, a part of the barrel is omitted for clarity. ]{\includegraphics[width=0.3\linewidth]{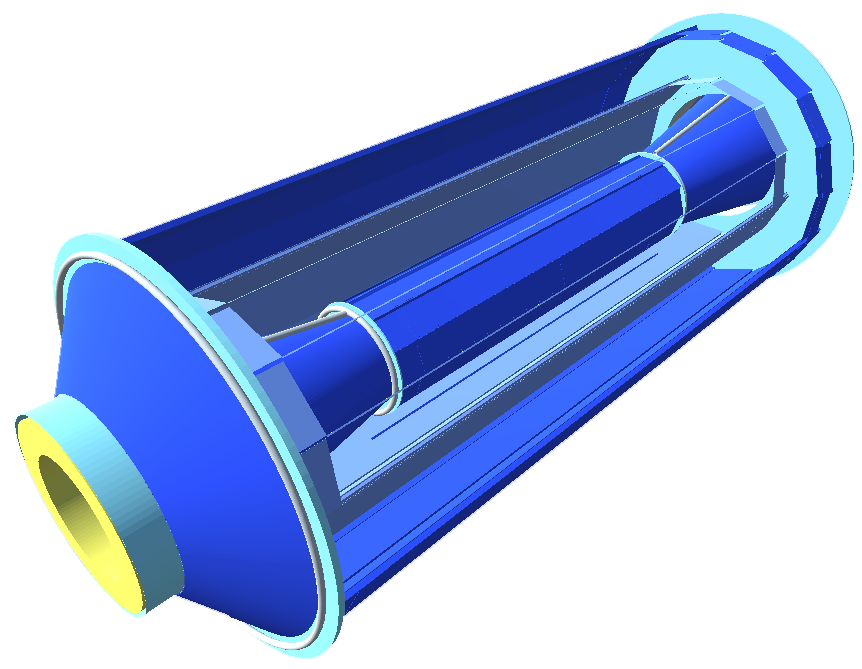}
    \label{fig:CEPC_vxd}}
    \caption{Some sub-detectors of CEPC shown in Unity. Subfigures (a) to (f) follow the detector's geometrical layout from the outermost to innermost components. }
    \label{fig:CEPC_sub}
\end{figure}

The X-Y cross-section view and Z-R longitudinal views of the CEPC detector in Unity, as depicted in Figure~\ref{fig:CEPC_cross}, allow for easy observation of the detector's intricate structures from the innermost to the outermost components, such as the cyan-green MUD. 
Additionally, specific sub-detectors and even detailed insights into the internal structure of sub-detectors can be examined individually, as shown in Figure~\ref{fig:CEPC_sub}. 
Among them, Figure~\ref{fig:CEPC_vxd} presents a partial structure of the innermost VXD, clearly showing its layered detector configuration and some supporting structures, with part of them omitted for clarity. 

\subsection{Performance}
\label{sec:performance}
The performance of the FBXWriter is evaluated for the aforementioned four detectors. 
All the conversions were performed on a computer with an Intel Core i7-13700H CPU and 64-GB memory. 
The time and memory required for the conversion process vary across detectors, primarily determined by the number of detector elements and nodes, and the precision of polygonal subdivision for curved surfaces. 
For the four detectors, the conversion time, number of polygons, and size of the FBX files are listed in Table~\ref{tab:performance}. 

\begin{table}
    \caption{The time consumption in detector conversion, number of polygons and size of the generated FBX files for the four detectors.}
    \centering
    \begin{tabular}{ccccccc}
    \hline
    \hline
      \hspace{1pt}
      Detector   & Conversion time~(s) & Number of polygons & FBX file size~(MB) \\ \hline
      \hspace{1pt}
       JUNO  & 80 & 51204 & 69.8 \\
       EicC & $\sim$0.2 & 30 & 1.6 \\
       BESIII & 27 & 23215 & 70.3 \\
       CEPC & 378 & 33868 & 85.7 \\
   \hline
   \hline
    \end{tabular}
    \label{tab:performance}
\end{table}

\section{Advantages and Further applications}
\label{sec:advantages}

\subsection{Advantages}
The method enables consistent and automated conversion of four commonly used detector descriptions from HEP software to FBX format, establishing a bridge between the intricate detectors of HEP experiments and modern industrial visualization technologies. 
Compared to the traditional HEP visualization techniques, industrial visualization offers advanced capabilities, long-term technical support, and multi-platform compatibility, significantly broadening the possibilities for HEP visualization. 

Although several previous efforts have explored feasible HEP visualization solutions, such as directly building detectors in Unity for JUNO event display software ELAINA~\cite{JUNO_unity}, using the GDML-FreeCAD-Pixyz-FBX workflow for BESIII visualization in Unity~\cite{BESIII_unity}, or leveraging the DD4hep-FBX interface for visualizations in industrial software~\cite{DD4hep_unity}, our method stands out in a few aspects. 
The following points summarize the advantages of this method compared to previous works: 
\begin{itemize}
    \item \emph{Comprehensive and universal.} 
    The proposed method supports the conversion of four commonly used detector description formats, covering 
    the majority of 
    detector construction methods in the offline software of HEP experiments. 
    This versatility allows the interface to seamlessly convert not only the existing detectors but also future detector designs into the FBX format, facilitating their visualization on industrial platforms. 
    Additionally, it aligns with the common community-defined format advocated by HSF~\cite{HSF}, offering a potential pathway towards a unified format for detector description. 
    \item \emph{Simplified steps and enhanced performance.}
    Since this method streamlines the entire process, requiring only one step to convert HEP detector geometry into FBX format, it significantly reduces the complexity of the conversion workflow, providing an efficient and effective approach to handling more complex detectors.
    \item \emph{Flexible settings.}
    Users can adjust parameters within the Geant4 that define shape descriptions, allowing for customization of the FBX rendering and conversion complexity. 
    This adds a high level of flexibility to the interface, ensuring its adaptability to different visualization needs. 
    \item \emph{Cost-effective.}
    Both the Geant4 interface and Unity platform are free for non-profit use. 
    This makes it possible to achieve comprehensive visualizations and further feature development with minimal cost, ensuring accessibility for a wide range of HEP experiments and applications. 
    \item \emph{Extensive shape support.}
    Geant4 offers a wide range of basic geometric shapes as well as complex geometries formed through Boolean operations, enabling the accurate representation and conversion of most shapes without encountering issues of unsupported geometries. 
\end{itemize}

In summary, this method presents undeniable advantages and occupies a unique position among these approaches, offering a streamlined and efficient process for integrating HEP detectors into cutting-edge visualization platforms. 

\subsection{Further application development}
\label{sec:further}
The visualization of detectors serves as the foundation for all visual representation functionalities. 
The conversion from the four specific detector descriptions from HEP software to FBX is unidirectional but not reversible, due to their different geometry construction systems.
However, since the conversion is automatic, once the detector geometry is modified, it is convenient to convert the updated geometry into a fresh FBX file. This feature is especially useful at the detector design stage of an experiment, when the detector designs change frequently or multiple versions of detector design exist in parallel.

\begin{figure}
    \centering
    \includegraphics[width=0.8\linewidth]{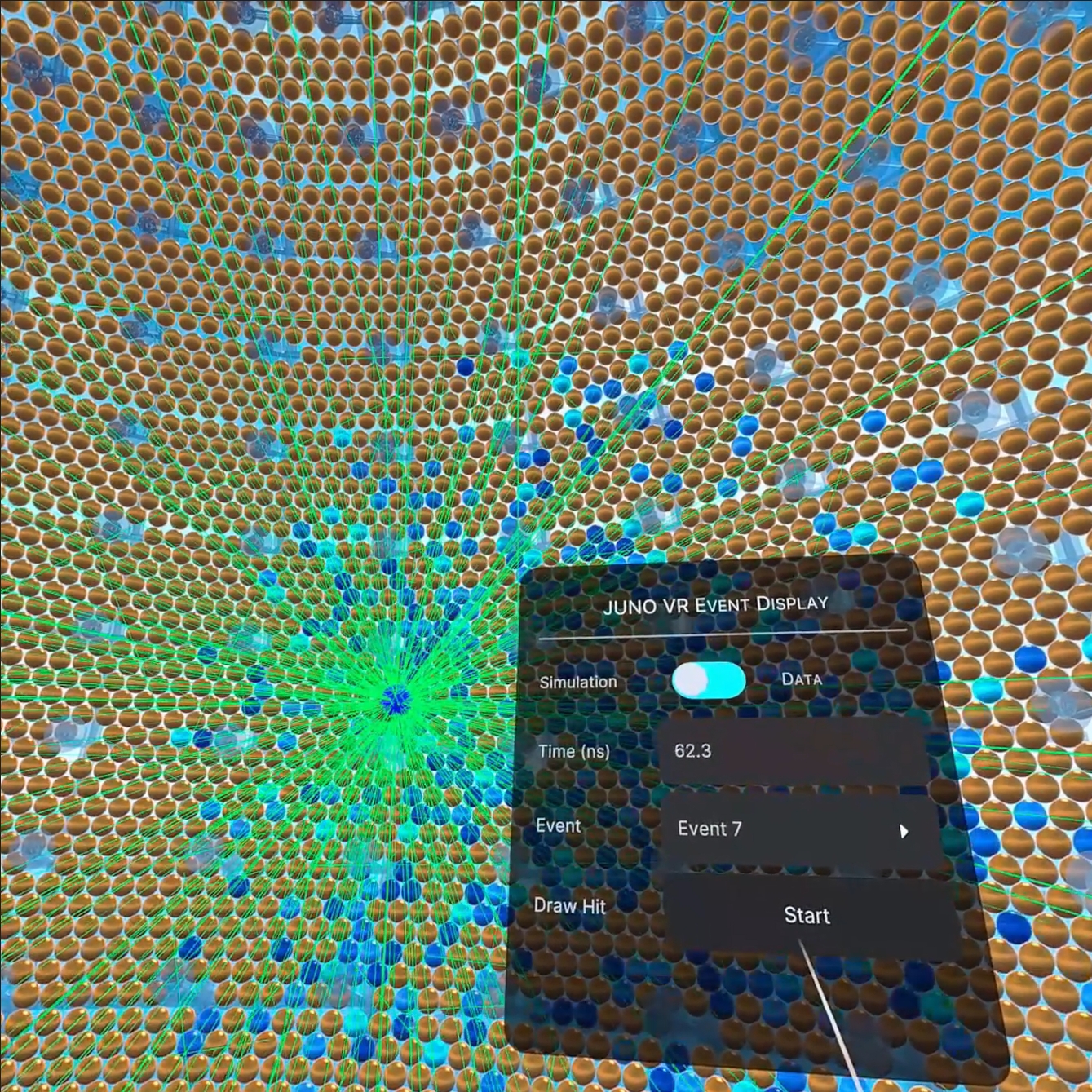}\\
    \caption{Display of a neutrino event in the preliminary design of the JUNO VR project. The yellow spheres are 20-inch PMTs with no hits, the light blue and dark blue spheres are 20-inch PMTs with photon detected in this event, and the green lines represent the propagation of photons in the liquid scintillator. The black panel is used for interactive operation with the VR touch controllers.}
    \label{fig:JUNO_VR_demo}
\end{figure}

Once the descriptions of HEP detectors are converted into FBX format, these FBX files can be directly imported into industrial 3D software such as Unity. 
This capability facilitates the creation of advanced applications, such as event display tools for data analyses in HEP experiments, as well as immersive VR~\cite{VR} or AR programs~\cite{AR} for education and outreach. 
Figure~\ref{fig:JUNO_VR_demo} shows an example of the VR application in the JUNO experiment. The detector geometry is automatically modeled with the detector description converted from Geant4 to FBX, as described in Section~\ref{sec:application}. 

Furthermore, the Unity platform allows for continuous enhancement of these applications. 
By integrating detector visualization with functional development, we achieve a seamless approach that facilitates the maintenance of detector consistency and allows for the efficient execution of comprehensive updates across all functionalities through a single command during the operation, maintenance, and upgrades of the HEP experiments. 
This integration enhances operational efficiency while enriching the overall user experience. 

\section{Summary}
\label{sec:summary}
In this study, we present a method to convert HEP detector descriptions from four major formats in HEP software to the FBX format, facilitating advanced visualization using industry tools, such as Unity. 
By utilizing the GDML and Geant4 software, this method leverages the existing FBXWriter interface in HSF to streamline the conversion process while maintaining strict consistency across the detector models, significantly enhancing both efficiency and flexibility in visualization. 
This advancement makes it more feasible to adopt cutting-edge visualization technologies in HEP experiments, with potential applications extending beyond basic detector visualization to include event display, VR, and AR developments. 

\section*{Acknowledgments}

We thank the helpful discussions in the offline software group of BESIII and JUNO collaboration, and thank the colleagues from IHEP, CAS working on CEPC, and colleagues from IMP, CAS working on EicC, for providing the detector description data files. 
This work is supported by the National Natural Science Foundation of China (Grant Nos. 12175321, W2443004, 11975021, 11675275, and U1932101), National Key Research and Development Program of China (Nos. 2023YFA1606000 and 2020YFA0406400), Strategic Priority Research Program of the Chinese Academy of Sciences (No. XDA10010900), National College Students Science and Technology Innovation Project, and Undergraduate Base Scientific Research Project of Sun Yat-sen University.




\end{document}